\documentclass[aps,prc,twocolumn,groupedaddress,showpacs,%
nofootinbib,showkeys,amsmath,floatfix]{revtex4}
\usepackage{graphics}
\usepackage{ifthen}
\begin{document}
\title{Description of nuclear octupole and quadrupole deformation
close to the axial symmetry and phase transitions in the octupole mode}
\author{P.G. Bizzeti}
\author{A.M. Bizzeti-Sona}
\affiliation{Dipartimento di Fisica, Universit\'a di Firenze\\
I.N.F.N., Sezione di Firenze\\
Via G. Sansone 1, 50019 Sesto Fiorentino (Firenze), Italy}
\date{\today}

\begin{abstract}
The dynamics of nuclear collective motion is investigated in the case
of reflection-asymmetric shapes.
The model is based on a
 new parameterization of the octupole and quadrupole degrees of
freedom, valid for nuclei close to the axial symmetry.
Amplitudes of oscillation in other degrees of freedom different from
the axial ones are assumed to be small, but not frozen to zero.
The case of nuclei which already possess a permanent quadrupole
deformation is discussed in some more detail  and  a simple solution is 
obtained at the critical point of the phase transition 
 between harmonic octupole oscillation and a permanent asymmetric shape.
The results are compared with experimental data of the Thorium 
isotopic chain. The isotope $^{226}$Th is found to be close to
the critical point.
\end{abstract}

\maketitle

\section{\label{S:1} Introduction}

Phase transitions in nuclear shapes have been recently observed 
at the boundary between regions characterized by different intrinsic
shapes of quadrupole deformation.
It has been shown by Iachello~\cite{iac1,iac2,iac2b}
that new dynamic symmetries, called E(5), X(5) and Y(5) hold, 
respectively, at the critical point
between spherical shape and $\gamma$--unstable deformation, 
between spherical and deformed 
axially symmetric shape and between deformed axial and triaxial shape.
Here, we are mainly interested in the second case. 
First examples of X(5)
symmetry in transitional nuclei have been found 
in $^{152}$Sm  \cite{casten1} and $^{150}$Nd
\cite{krue1}. 
Other candidates for the X(5)
symmetry in different nuclear regions have been reported
later~\cite{biz1,hutt,clar,fran,bizoct2,bijk03,bijk04,tone}.
From the theoretical point of view, slightly different potentials have
been explored by Caprio~\cite{capr04} and by Bonatsos 
\textit{et al.}~\cite{bonat04}, while the evolution of the $s=2$ band 
when the lower border of the square well potential is displaced from
zero has been investigated by Pietralla and Gorbachenko~\cite{piet04}.

A similar phase transition (\textit{i.e.}, 
from shape oscillation to
permanent deformation, conserving the axial symmetry of the system)
could take place also in the octupole degree of freedom. We have 
found a possible example of such a phase transition 
in the Thorium isotope chain, with the critical
point close to the mass 226. 
In this case, the octupole mode is combined with a stable
quadrupole deformation. 
Preliminary results have been reported in
two recent Conferences~\cite{bizoct1,bizoct2}. 

In order to provide a theoretical frame where to discuss the different
aspects of the octupole motion, we introduce a new
parameterization of the quadrupole and octupole degrees of freedom,
valid in conditions close to the axial symmetry. In this limit, a model
similar to the classical one by Bohr~\cite{bohr52} has been developed.
This is the subject of the first part of this paper.
In the second part, the model is used to discuss the evolution of the
octupole mode along the isotopic chain of Thorium, and the results are
compared with the experimental data.

\section{\label{S:2} Theoretical frame for combined 
quadrupole and 
octupole excitations}
\subsection{\label{S:2.1} Previous investigations of the Octupole 
plus Quadrupole deformation}
Reflection--asymmetric nuclear shapes have been discussed in a number
of papers, either in terms of surface Quadrupole + Octupole
deformation (Bohr geometrical approach) or with an extended
Interacting Boson Model (algebraic approach). The latter,
 proposed in 1985 by Engel and 
Iachello~\cite{iac3}, has been recently used by Alonso \textit{ et
al.}~\cite{alonso}, Raduta \textit{et al.}~\cite{radu03,radu03b}, 
Zamfir and Kusnezov~\cite{zam01,zam03}. 
An alternative approach assuming $\alpha$--cluster configurations
has been discussed by Shneidman \textit{et al.}~\cite{shnei}.
In the frame of the geometrical approach, a
number of theoretical investigations of the octupole vibrations 
around a stable quadrupole deformation have been reported in the
last 50 years~\cite{roho,butler,lipas,denis,jolos,minkov,minkov04}.
 Most of them, however, are limited to the case of axial symmetry. 
This approach has been criticized, \textit{e.g.}, by Donner and
Greiner~\cite{donner}, who have stressed the fact that all terms of a
given tensor order must be taken into account for a consistent
treatment.
To do this,  Donner and Greiner renounce to the use of an ``intrinsic
frame'' referred to the principal axes of the overall tensor of
inertia and choose to define the octupole amplitudes in the
``intrinsic frame'' of the quadrupole mode alone. 
In this approach, definite predictions have been obtained at the limit
where the octupole deformations are ``small'' in comparison with the
quadrupole ones~\cite{eisen1}.
\nopagebreak
\subsection{\label{S:2.2} The new parameterization}
Here we adopt a different approach, which can be useful also in the
case of comparable octupole and quadrupole deformation, close to the
axial--symmetry limit. Namely, we
choose as ``intrinsic'' reference frame the principal axes of
the overall tensor of inertia, as it results from the combined 
quadrupole and octupole deformation. 
The definitions of quadrupole and octupole amplitudes 
$a^{(\lambda ) }_\mu$,
with $\lambda =2,3$ and 
$a^{(\lambda ) }_{-\mu}=(-)^\mu \ {a^{(\lambda ) }_\mu}^*$,
are recalled in the Appendix~\ref{S:a1}. All these amplitudes are
defined in the (non inertial) intrinsic frame. To this purpose, in the
case of the quadrupole mode alone, it is 
enough to assume
$a^{(2)}_2\!=a\!^{(2)}_{-2}$ real and $a^{(2)}_{\pm 1}\!=\!0$, with
the standard parameterization in terms of $\beta _2$ and $\gamma _2$:
\begin{subequations}
\label{E:1}
\begin{eqnarray}
a^{(2)}_0 &=& \beta_2 \cos \gamma_2  \\[1pt]
a^{(2)}_1 &=& 0  \\[1pt]
a^{(2)}_2 &=& \sqrt{1/2}\ \beta_2\ \sin \gamma_2\ .
\end{eqnarray}
\end{subequations}
For the octupole mode
alone, a  parameterization suitable to this purpose has been proposed
in 1999 by Wexler and Dussel~\cite{wexler}. We adopt here a very
similar one: 
\begin{subequations}
\label{E:2}
\begin{eqnarray}
a^{(3)}_0 &=& \beta_3 \ \cos \gamma_3 
 \\[1pt]
a^{(3)}_1 &=& -  (5/2)\ \left( X + i Y \right)\ \sin \gamma_3\\
a^{(3)}_2 &=& \sqrt{1/2}\ \beta_3\ \sin \gamma _3  \\[1pt]
a^{(3)}_3 &=& X \left[ \cos \gamma_3 + (\sqrt{15}/2)\ \sin \gamma_3
\right] \nonumber \\[1pt]
  &+& i\ Y \left[ \cos \gamma_3 - (\sqrt{15}/2)\ \sin \gamma_3
\right] \ .
\end{eqnarray}
\end{subequations}
With this choice, the tensor of inertia turns out to be diagonal (see
Appendix~\ref{S:a1}). 

In both cases, one has to consider, in addition to the intrinsic
variables ($\beta_2,\ \gamma_2$ for the quadrupole, or 
$\beta_3,\ \gamma_3\ ,X\ ,Y$ for the octupole), the three Euler angles
defining the orientation of the intrinsic frame in the laboratory
frame, in order to reach a number of parameters equal to the number of
degrees of freedom (5 for the quadrupole, 7 for the octupole).

Unfortunately, the situation is not so simple when quadrupole and
octupole modes are considered together, as the intrinsic frames of the
two modes do not necessarily coincide.
We shall limit our discussion to situations close to the axial
symmetry limit -- in which, obviously, the two frames coincide -- and
define a parametrization which automatically sets to zero the three
products of inertia $\mathcal{J}_{\kappa, \kappa ^\prime}$ 
($\kappa \neq \kappa ^\prime$) \textsl{ up to the first order in the
amplitudes of non--axial modes}.

To this purpose we put 
\begin{equation}
a^{(\lambda )}_\mu = \bar{a}^{(\lambda )}_\mu + 
\tilde{a}^{(\lambda )}_\mu 
\label{E:3}
\end{equation}
where $\bar{a}^{(\lambda )}_\mu$ are defined according to the 
eq.s~(\ref{E:1},\ref{E:2}) and $\tilde{a}^{(\lambda )}_\mu$ are
correction terms, which are assumed to be small compared to the axial
amplitudes $a^{(\lambda)}_0$, but of the same order of magnitude as the
other non-axial terms.
It will be enough to consider these corrections only
for those amplitudes which, according to eq.s~(\ref{E:1},\ref{E:2}),
 are either  zero or small of the second order: the imaginary part 
of $a^{(2)}_2,\ a^{(3)}_2$ and the 
real and imaginary parts of $a^{(2)}_1,\ a^{(3)}_1$.
The six ``new'' first--order amplitudes added to those of
eq.s~(\ref{E:1},\ref{E:2}) are, however, not independent from one
another, if we choose as the reference systeme the one in which 
the three 
products of inertia turn out to be zero.

The expressions of the inertia tensor as a function of the
deformation parameters, obtained with the Bohr
assumptions~\cite{bohr52} of not--too--big deformations and
irrotational flow, are given in the Appendix~\ref{S:a1}. In order to
simplify the notations, from now on we consider the inertia parameters
$B_2,\ B_3$ included in our definitions of the amplitudes $a^{(\lambda
) }_\mu$, which therefore correspond to 
$\sqrt{B_\lambda}\ a^{(\lambda ) }_\mu$ in the original Bohr notations.
From the eq.s~(\ref{E:a.10}), and retaining only terms 
of the first order
in the small amplitudes $ a^{(\lambda ) }_\mu$ with $\mu \neq 0$, we
obtain the conditions
\begin{subequations}
\label{E:3a}
\begin{eqnarray}
\mathcal{J}_{12}&=& - 2\sqrt{6}\left( \beta_2 {\rm Im}\ \tilde{a}^{(2)}_2+
\sqrt{5}\beta_3 {\rm Im}\ \tilde{a}^{(3)}_2 \right) = 0\\*
\mathcal{J}_{13}&+&i \mathcal{J}_{23}\ = 
\ \sqrt{6}\ \left( \beta_2 \tilde{a}^{(2)}_1+
\sqrt{2}\beta_3 \tilde{a}^{(3)}_1 \right) = 0
\end{eqnarray}
\end{subequations}
 which are satisfied (at the leading order) if we put
\begin{equation}
\begin{array}[c]{lp{0.1mm}lp{0mm}l}
\tilde{a}^{(2)}_1=\!\frac{-\sqrt{2}\beta_3}{\sqrt{\beta_2^2+2\beta_3^2}}
\left(\eta + i \zeta \right) && 
\tilde{a}^{(3)}_1=\!\frac{\beta_2}{\sqrt{\beta_2^2+2\beta_3^2}}
\left(\eta + i\zeta \right) \\[8pt] 
{\rm Im}\;\tilde{a}^{(2)}_2=\!\frac{-\sqrt{5}\beta_3}
{\sqrt{\beta_2^2+5\beta_3^2}}
\xi &&
{\rm Im}\;\tilde{a}^{(3)}_2=\!\frac{\beta_2}{\sqrt{\beta_2^2+5\beta_3^2}}
\xi 
\end{array}
\label{E:4}
\end{equation}
with the new parameters $\eta,\ \zeta$ and $\xi$ small of the first
order.
It is clear that only the ratios of the relevant amplitudes are
constrained by the eq.s (\ref{E:3a}). The definition of the new variables
given in each line of eq.s (\ref{E:4}) contain therefore an arbitrary
factor. Our choice (and in particular for the square--root factors at 
the denominators) has some distinguished advantage, which will result
clear from the classical expression of the kinetic energy, discussed
in the next paragraph.

In the intrinsic reference frame, and at the same 
order of approximation, the 
values of the three principal moments of inertia can be derived from 
eq.s~(\ref{E:a.8},\ref{E:a.9}):
\begin{subequations}
\begin{eqnarray}
\mathcal{J}_1 &=& 3(\beta_2^2 + 2\beta_3^2) +2\sqrt{3}(\beta_2^2
\gamma_2 +\sqrt{5} \beta_3^2 \gamma_3) \\*[3pt]
\mathcal{J}_2 &=& 3(\beta_2^2 + 2\beta_3^2) -2\sqrt{3}(\beta_2^2
\gamma_2 + \sqrt{5} \beta_3^2 \gamma_3) \\*[3pt]
\mathcal{J}_3 &=& 4(\beta_2^2 \gamma_2^2 +\beta_3^2 \gamma_3^2)
+18(X^2 + Y^2) \nonumber \\*[3pt]
&& + 2 (\eta^2 + \zeta^2) 
+8 \xi^2\ .
\end{eqnarray}
\end{subequations}
With the amplitudes given by eq.s (\ref{E:1},\ref{E:2}) the principal
axes of the quadrupole would coincide with those of the octupole.
It is not necessarily so with our more general assumptions.
 When $a^{(\lambda)}_1\neq 0$, the
axis 3 of the tensor of inertia for the quadrupole mode alone does not
coincide with that of the octupole. If $a^{(\lambda)}_1= 0$, but 
Im$\ a^{(\lambda)}_2\neq 0$, the misalignment concerns the other two
principal axes perpendicular to the common axis 3.

\begingroup
\ifthenelse{\linewidth = \textwidth}{\begin{turnpage} }{}
\squeezetable
{\renewcommand{\arraystretch}{1.30}
\begin{table*}[t]
\caption{\label{T:1} The  matrix of inertia $\mathcal{ G}$: 
leading terms and relevant first-order terms.
Other first-order terms are indicated with
the symbol $\approx 0$. As the matrix is symmetric, first-order terms
in the last three columns are not explicitly shown. 
Here $\mathcal{ J}_1= 3(\beta_2^2 + 2\beta_3^2) +2\sqrt{3}(\beta_2^2
\gamma_2 + \sqrt{5} \beta_3^2 \gamma_3)$; 
$\mathcal{ J}_2 = 3(\beta_2^2 + 2\beta_3^2) -2\sqrt{3}(\beta_2^2
\gamma_2 + \sqrt{5} \beta_3^2 \gamma_3)$
; and
$\mathcal{ J}_3 = 4(\beta_2^2 \gamma_2^2 +\beta_3^2 \gamma_3^2)
+18(X^2 + Y^2) + 2 (\eta^2 + \zeta^2) 
+8 \xi^2$.\\
The determinant of the matrix is 
\ ${G = {\rm Det}\mathcal{ G} =
1152 \beta_2^2 \beta_3^2 \big( \beta_2^2 + 2 \beta_3^2 \big) ^2
\big( \beta_2^2 + 5 \beta_3^2 \big) ^{-1}
\big( \beta_2^2 \gamma_2 + \sqrt{5}\ \beta_3^2 \gamma_3 \big) ^2}$.
}
\begin{ruledtabular}
\begin{tabular}{l|ccccccccc|ccc}
& 
$\dot{\beta_2}$ & 
$\dot{\gamma_2}$ & 
$\dot{\beta_3}$ & 
$\dot{\gamma_3}$ & 
$\dot{X}$ & 
$\dot{Y}$ & 
$\dot{\xi} $ & 
$\dot{\eta}$ &  
$\dot{\zeta}$ & 
$q_1$ & 
$q_2$ & 
$q_3$ \\
\hline
$\dot{\beta_2}$ & 1 & 0  & 0  & 0  & 0  & 0  & 
0 & 0 & 0 & $\left[ ..\right]  $ & 
$\left[ ..\right] $ & 0\\
$\dot{\gamma_2}$ & 0  & 
$\beta_2^2$
 &0 & 0  & 0  & 0  & 0  & 0  & 0 & 
$\left[ ..\right]$ & 
$\left[ ..\right]$ &
 $ (..)$ \\
$\dot{\beta}_3$ & 0 & 0 & 1 &
0  & 0 & 0  & 0 & 0 & 0 &
$\left[ ..\right]$ &
$\left[ ..\right]$ &
 0\\
$\dot{\gamma_3}$ & 0  & 0  & 0 &
 $\beta_3^2$  & 
$\left[ \sqrt{15}X\right] $   & 
$\left[ -\!\sqrt{15}Y\right]$  &  0  & 
$\left[ -\frac{5\beta_2 X}{\sqrt{\beta_2^2 
+ 2\beta_3^2}}\right]$ & 
$\left[ -\frac{5\beta_2 Y}{
\sqrt{\beta_2^2+ 2\beta_3^2}}  \right]$  &  
$\left[ ..\right]$ &
$\left[ ..\right]$ &
 $(..)$ \\[6pt]
$\dot{X}$ & 0  & 0  & 0  & 
$\left[ \sqrt{15}X\right] $ & 
$2\!+\!2\sqrt{15} \gamma_3$  & 0  &  0  &
$\left[ -\frac{5\beta_2 \gamma_3}{
\sqrt{\beta_2^2+ 2\beta_3^2}} \right] $  & 0  &  
$\left[ ..\right] $ & 
$\left[ .. \right]$ &  
$(..)$ \\
$\dot{Y}$ & 0  & 0  & 0  & 
$\left[ -\!\sqrt{15}Y\right]$ 
 & 0  & $2\!-\!2\sqrt{15} \gamma_3$ &  0  & 0  &
$\left[ -\frac{5\beta_2 \gamma_3}{
\sqrt{\beta_2^2+ 2\beta_3^2}} \right] $  &  
$\left[ .. \right]$ &  
$\left[ ..\right] $ &
 $(..)$ \\
$\dot{\xi}$ 
& 0 & 0  & 0 & 
 0  & 0  & 0  &  
 2 & 0  & 0  &  
$\left[ ..\right]$  &
$\left[ .. \right]$  &
 $( ..)$ \\
$\dot{\eta}$ & 
 0 &  0  & 0 &
$\left[ -\frac{5\beta_2 X}{
\sqrt{\beta_2^2+ 2\beta_3^2}} \right] $  & 
$\left[ -\frac{5\beta_2 \gamma_3}{
\sqrt{\beta_2^2+ 2\beta_3^2}} \right] $ &
 0  & 0 & 2 & 0  & 
$\left[ ..\right]$  & 
$\left[ ..\right]$ &
 $(.. )$ \\[6pt]
$\dot{\zeta}$ & 
0 &  0  &  0 &
$\left[ -\frac{5\beta_2 Y}{
\sqrt{\beta_2^2+ 2\beta_3^2}} \right]$  & 0  &
$\left[ -\frac{5\beta_2 \gamma_3}{
\sqrt{\beta_2^2+ 2\beta_3^2}} \right]$  &  0  & 0  &
 2 &
$\left[ ..\right] $ &
$\left[ ..\right] $ & 
$(..)$\\[6pt]
\hline
$q_1$ & $\approx 0$  & $\approx 0$  & $\approx 0$  & $\approx 0$  
& $\approx 0$  & $\approx 0$  & $\approx 0$  & $\approx 0$  
& $\approx 0$ &
 ${\mathcal{ J}_1}$  & 0  &  0 \\
$q_2$  & $\approx 0$  & $\approx 0$  & $\approx 0$  & $\approx 0$ 
 & $\approx 0$  & $\approx 0$  & $\approx 0$  & $\approx 0$  
& $\approx 0$  
& 0 & 
${\mathcal{ J}_2}$  &
0 \\
$q_3$ & 0  & 
 $ \frac{-\!\sqrt{40}\beta_2 \beta_3 \xi
}{\sqrt{\beta_2^2+5\beta_3^2}}$ & 0  &
 $\frac{\sqrt{8}\beta_2 \beta_3 \xi
}{\sqrt{\beta_2^2+5\beta_3^2}}$  & 
 $6Y$  & 
 $-\!6X$  & 
 $\frac{\sqrt{8}\beta_2 \beta_3 (\sqrt{5}\gamma_2\!-\!\gamma_3 )
}{\sqrt{\beta_2^2+5\beta_3^2}}\phantom{.}$ &  
 $2 \zeta $\phantom{.}&  
 $-2 \eta $  & 
0 & 0 & 
 $\mathcal{ J}_3$ \\[6pt]
\end{tabular}
\end{ruledtabular}
\end{table*}
}
\ifthenelse{\linewidth =\textwidth}{\end{turnpage} }{}
\endgroup
\subsection{\label{S2:3} The classical expression of the kinetic energy}
Now it is possible to express the classical kinetic energy (as given
by Bohr hydrodynamical model) in terms of the new variables and
of the intrinsic components $q_\kappa $
of the angular velocity. The classical expression has the form
\begin{equation}
T=\frac{1}{2}\sum \dot{Q}_\mu \mathcal{G}_{\mu \nu} \dot{Q}_\nu
\end{equation}
where $\dot{Q} \equiv \{\dot{\xi}_1,\dot{\xi}_2, ...,
\dot{\xi}_9,q_1,q_2,q_3\}$, $\dot{\xi}_\mu \ (\mu =1,...,9)$ are the
time derivative of the nine parameters we have just defined, and
$q_1,\ q_2,\ q_3$ are the intrinsic components of the angular velocity
of the intrinsic system with respect to an inertial frame.
The elements of the matrix $\mathcal G$ 
(leading terms and 
relevant first-order terms) are shown in the Table~\ref{T:1}. 

The determinant $G = {\rm Det}\ \mathcal{G}$ takes the form
\[
G\propto \beta_2^2 \beta_3^2 \big( \beta_2^2 + 2 \beta_3^2 \big) ^2
\big( \beta_2^2 + 5 \beta_3^2 \big) ^{-1}
\big( \beta_2^2 \gamma_2 + \sqrt{5}\ \beta_3^2 \gamma_3 \big) ^2
\]
and, at the limit $\beta_3 \ll \beta_2$, results to be proportional 
to $\beta_2^8$, and therefore
consistent with that of the Bohr model for a pure quadrupole motion.
This is a consequence of our choice of the normalization factors in
the eq.s (\ref{E:4}). This choice has other advantages: all the non diagonal
terms involving the time derivatives of $\beta_2$ or $\beta_3$ and 
 either the derivative of one
of the other intrinsic amplitudes or $q_3$ turn out to be zero in the
present approximation. Other non diagonal elements are small (of the
first order) in the ``small'' amplitudes $\gamma_2,\ \gamma_3,\ X,\ Y,
\ \xi,\ \eta,\ \zeta$. In situations close to the axial symmetry, they 
 have negligible effect on the results (see Appendix~\ref{S:a2}), 
with the only exception of
 elements of the
last line and column. The latter, in fact, are still small of the
first order, but must be compared with the diagonal element ${\mathcal
J}_3$,
which is small of the second order in the ``small'' non--axial 
amplitudes.
These terms play an important role in the treatment of the intrinsic
component of the angular momentum along the approximate
axial--symmetry axis, which will be discussed in the next paragraph.

\subsection{\label{S:2.4} Intrinsic components of the angular momentum}

According to the classical mechanics, the components $L_1$, $L_2$ and
$L_3$ of the angular momentum in the intrinsic frame
are obtained as the derivatives of the total kinetic energy 
with respect to the corresponding intrinsic component of the 
angular velocity:
\begin{equation}
L_k = \frac{\partial T}{\partial q_k} \ .
\label{E:6.1}
\end{equation}
The part of the kinetic energy depending on the component $q_k$ 
has the form
\begin{equation}
T_k(q_k) = \frac{1}{2} \mathcal{J}_k q_k^2 + F_k q_k
\label{E:6.2}
\end{equation}
where $F_k$ is a function of the dynamical variables $\xi_i$ 
and of their
derivatives with respect to the time, and is 
{\em small of the first order}
according to our definition. As for the moments of inertia, 
$\mathcal{J}_3$ is 
 {\em small of the second order}, while $\mathcal{J}_1$, 
$\mathcal{J}_2$ are not small. 
According to eq.(\ref{E:6.1}), we have

\begin{eqnarray}
L_k &=&  \mathcal{J}_k q_k + F_k  \nonumber \\[1pt]
q_k &=& \frac{ L_k - F_k }{ \mathcal{J}_k }  
\label{E:6.3a}
\end{eqnarray}
\begin{eqnarray}
T_k(L_k) &=& \frac{1}{2}  \mathcal{J}_k 
\  \left[ \frac{1}{ \mathcal{J}_k }
\big( L_k - F_k \big) \right] ^2 + F_k \left[ \frac{1}{ \mathcal{J}_k }
\big( L_k - F_k \big) \right]  \nonumber \\[1pt]
&=& \frac{L_k^2}{2  \mathcal{J}_k } - \frac{F_k^2}{2 \mathcal{J}_k } 
\ .
\label{E:6.3}
\end{eqnarray}
For $k=1,2$ the second term is small of the second order and can 
be neglected.
It is not so for $k=3$, as $\mathcal{J}_3$ is also small and of 
the same order as $F_k^2$.
In more detail, we have
\begin{eqnarray}
L_3 &=& \mathcal{J}_3 q_3 +
\Big[ \ \frac {\sqrt{8}\ \beta_2 \beta_3}{\sqrt{\beta_2^2 + 5 \beta_3^2}}
 \big( \gamma \dot{\xi} - \xi \dot{\gamma} \big) \nonumber \\*
&+& 6 \big( Y \dot{X} - X\dot{Y} \big)
+ 2 \big(  \zeta \dot{\eta} -  \eta \dot{\zeta}
\big) \ \Big] 
\label{E:4.1}
\end{eqnarray}
where we have put
\begin{equation}
\gamma = \sqrt{5}\ \gamma_2 - \gamma_3\ .
\label{E:4.1a}
\end{equation}
At this point, it will be convenient to express the variables
$\gamma_2$ and $\gamma_3$ as linear combinations of two new variables,
one of which is, obviously, $\gamma = \sqrt{5} \gamma_2 - \gamma_3$.
The other one, that we call $\gamma_0$,  can be chosen proportional to
the linear combination which enters in the expression of the
determinant $G$, 
\begin{equation}
\gamma_0 = c \left(\beta_2^2 \gamma_2 + \sqrt{5}\ \beta_3^2 \gamma_3 
\right).
\label{E:4.2}
\end{equation}
With this choice, we obtain
\begin{eqnarray}
\gamma_2 &=& \frac{ \gamma_0/c + \sqrt{5}\ \beta_3^2 \gamma }
{\beta_2^2 + 5 \beta_3^2 }  \nonumber \\[1pt]
\gamma_3 &=& \frac{\sqrt{5}\ \gamma_0/c - \beta_2^2 \gamma }
{\beta_2^2 + 5 \beta_3^2}
\label{E:4.3}
\end{eqnarray}
and,  at the leading order,
\begin{equation}
\beta_2^2 \dot{\gamma}_2^2 +\beta_3^2 \dot{\gamma}_3^2
= \frac{\beta_2^2\beta_3^2}{\beta_2^2 + 5 \beta_3^2}\dot{\gamma}^2
+ \frac{1}{ \beta_2^2 + 5 \beta_3^2 } \ ( \dot{\gamma}_0/c \big) ^2
\ .
\label{E:4.4}
\end{equation}
In deriving the eq.~\ref{E:4.4}, the factor $1/c$ has been considered
constant. We may note,however, that the same result holds \textsl{at
the leading order} also if $1/c$ is a function of $\beta_2$ and/or
$\beta_3$. In fact, terms involving the time derivative of $1/c$ also
contain the ``small'' quantity  $\gamma_0$, and their effect is
negligible in the present approximation (see Appendix \ref{S:a2}).
\textit{E.g.}, one could choose for $1/c$ a quadratic expression in
$\beta_2 ,\ \beta_3$, in order to obtain for $\gamma_0$ an
adimensional quantity, like $\gamma_2 ,\ \gamma_3$ and $\gamma$.
The same argument applies for possible redefinitions of other
``small'' variables, like $\gamma$.

The expression (\ref{E:4.1}) can be somewhat simplified with the 
substitutions
\begin{subequations}
\label{E:4.5}
\begin{eqnarray}
X &=& w \sin \vartheta \\*
Y &=& w \cos \vartheta \\*
\eta &=&  v \sin \varphi \\*
\zeta &=& v \cos \varphi \\*
\xi &=&   u \sin \chi  \\*[1pt]
\gamma &=& \sqrt{2}\ \left( \sqrt{\beta_2^2+5\beta_3^2 }\ /
\ \beta_2 \beta_3 \right) u \cos \chi \\*[1pt]
\frac{\gamma_0}{c} &=& f(\beta_2,\beta_3) \sqrt{\beta_2^2+5\beta_3^2 }
\ u_0
\end{eqnarray}
\end{subequations}
which gives for the determinant of the matrix $\mathcal G$ 
\begin{equation}
G={\rm Det}\;\mathcal{G}=2304 \big( \beta_2^2+2\beta_3^2 \big) ^2
\ u_0^2\ v^2 
u^2 w^2\  f^2(\beta_2, \beta_3).
\label{E:4.6}
\end{equation}

The choice of the function $f(\beta_2, \beta_3)$ is irrelevant for
what concerns the angular momentum. Non diagonal terms (small of the
first order) would depend on this choice, but their effect is
negligible (see Appendix~\ref{S:a2}).  As a criterion to define the
form of the function $f$, we observe that for permanent quadrupole
deformation $\beta_2=\bar{\beta}_2$ and at the limit $\beta_3^2 \ll
\beta_2^2$ our value of $G$ must agree with the result given, at this
limit, by Eisenberg and Greiner~\cite{eisen1}. This happens if the
function $f$ we have left undetermined tends to a constant when 
 $\beta_2 \rightarrow \bar{\beta}_2$. 
We adopt here the simplest possible choice,
\textit{i.e.} $f(\beta_2, \beta_3)=1$, to obtain the matrix  
$\mathcal{G}$
given in the Table \ref{T:2}, and (at the leading order)
\begin{table}[ht]
\caption{\label{T:2}  The  matrix of inertia $\mathcal{ G}$ 
after the introduction of the new variables $u_0$, $v$, $u$,
$w$,  $\varphi$,  $\chi$, and $\vartheta$ (see text).
Here, $\mathcal{ J}_1 = \mathcal{ J}_2 = 3(\beta_2^2 + 2\beta_3^2)$, and
$\mathcal{ J}_3 = 4 
\ u_0^2 + 
2v^2 + 8u^2 + 18w^2$.
Only the leading terms are shown. Neglected terms are small of the
first order (or smaller) in the sub-matrix involving only 
 $\dot{\beta_2}$, $\dot{\beta_3}$, $\dot{u_0}$, 
$\dot{v}$, $\dot{u}$, $\dot{w}$, 
 $q_1$ and $q_2$; of the third order (or smaller) in the sub-matrix 
involving only
$\dot{\varphi} $, $\dot{\chi}$,  $\dot{\vartheta}$ and $q_3$;
of the second order (or smaller) in the rest of the matrix.}
\begin{ruledtabular}
\begin{tabular}{l|ccccccccc|ccc}
& $\dot{\beta_2}$ & $\dot{\beta_3}$ & $\dot{u_0}$ & 
$\dot{v}$ & $\dot{u}$ & $\dot{w}$ 
& $\dot{\varphi} $ & $\dot{\chi}$ &  $\dot{\vartheta}$ & $q_1$
& $q_2$  & $q_3$\\
\hline
$\dot{\beta_2}$ & 1 & 0 & 0 & 0 & 0 & 0 & 0 & 0 & 0 & 0 & 0 & 0 \\
$\dot{\beta}_3$ & 0 & 1 & 0 & 0 & 0 & 0 & 0 & 0 & 0 & 0 & 0 & 0 \\
$\dot{u_0}$ & 0 & 0 & $1$ 
 & 0 & 0 & 0 & 0 & 0 & 0 & 0 & 0 & 0 \\
$\dot{v}$    & 0 & 0 & 0 & 2 & 0 & 0 & 0 & 0 & 0 & 0 & 0 & 0 \\
$\dot{u}$   & 0 & 0 & 0 & 0 & 2 & 0 & 0 & 0 & 0 & 0 & 0 & 0 \\
$\dot{w}$       & 0 & 0 & 0 & 0 & 0 & 2 & 0 & 0 & 0 & 0 & 0 & 0 \\
$\dot{\varphi}$ & 0 & 0 & 0 & 0 & 0 & 0 & $2v ^2$   & 0 & 0 
                                           & 0 & 0 &  $2v ^2$ \\
$\dot{\chi}$    & 0 & 0 & 0 & 0 & 0 & 0 & 0 & $2u^2$ & 0 
                                    & 0 & 0 &  $4u^2$ \\
$\dot{\vartheta}$ & 0 & 0 & 0 & 0 & 0 & 0 & 0 & 0 & $2 w^2$
                                           & 0 & 0 & $6 w^2$ \\
\hline
$q_1$ & 0 & 0 & 0 & 0 & 0 & 0 & 0 & 0 & 0 & $\mathcal{ J}_1$ & 0 & 0 \\
$q_2$ & 0 & 0 & 0 & 0 & 0 & 0 & 0 & 0 & 0 & 0 & $\mathcal{ J}_2$ & 0 \\
$q_3$ & 0 & 0 & 0 & 0 & 0 & 0 & $2v^2$ & $4u^2$ & $6w^2$
                                          & 0 & 0 & $\mathcal{ J}_3$ \\
\end{tabular}
\end{ruledtabular}
\end{table}
\begin{eqnarray}
G&=&{\rm Det}\;\mathcal{G}=2304 \big( \beta_2^2+2\beta_3^2 \big) ^2
\ u_0^2\ v^2 u^2 w^2 
\label{E:6.51a}
\\
\mathcal{J}_3 &=&4 u_0^2 + 2 v^2 + 8 u^2 + 18 w^2 
\label{E:6.51b}
\\
2T &=& \dot{\beta}_2^2 + \dot{\beta}_3^2 + \dot{u}_0^2 
\nonumber \\*
&+& 2  \big( \dot{v}^2 + v^2 \dot{\varphi}^2 \big) 
+ 2 \big(\dot{u}^2 +u^2 \dot{\chi}^2 \big) 
+ 2 \big( \dot{w}^2 + w^2 \dot{\vartheta}^2 \big) \nonumber  \\*[1pt]
&+& 2 q_3\left[ 2  v^2 \dot{\varphi} + 4u^2 \dot{\chi}
+ 6 w^2 \dot{\vartheta} \right] \nonumber \\*[1pt]
&+& \mathcal{J}_1 q_1^2 + \mathcal{J}_2 q_2^2 +\mathcal{J}_3 q_3^2\ . 
\label{E:6.51c}
\end{eqnarray}

\subsection{\label{S:2.5} Classification of elementary 
excitations with respect to $K^\pi$}

We can now deduce the intrinsic components of the angular momentum:
\begin{subequations}
\label{E:4.7}
\begin{eqnarray}
L_1 &=& \mathcal{J}_1 q_1 \\*
L_2 &=& \mathcal{J}_2 q_2 \\*
L_3 &=& \mathcal{J}_3 q_3 +
\left[  2  v ^2 \dot{\varphi}
+4 u^2 \dot{\chi} 
+6 w^2 \dot{\vartheta} 
\right] \nonumber \\*
&=&  2 v^2 \big( q_3 +\dot{\varphi} \big) 
+ 4 u^2 \big(2 q_3+ \dot{\chi} \big)
+ 6 w^2 \big( 3 q_3+ \dot{\vartheta} \big)\nonumber \\[1pt]
&\phantom{+}& + 4 u_0^2 q_3\ .  
\end{eqnarray}

In the same way, we can obtain the classical moments conjugate to
$\chi$, $\vartheta$ and $\varphi$
(we observe that none of these variables appears in the expressions of
$G$ or $\mathcal{J}_3$):
\begin{eqnarray}
p_\varphi &=& 2  v^2 ( \dot{\varphi} 
+ q_3 ) \\
p_\chi &=& 2 u^2
\left( \dot{\chi} + 2 q_3 \right) \\
p_\vartheta &=& 2 w^2 ( \dot{\vartheta} + 3 q_3 ) \ .
\end{eqnarray}
\end{subequations}
Now we can solve the system of equations 
(\ref{E:4.7}) with respect to
the variables $q_1,\ q_2,\ q_3,\ \dot{\varphi},\ \dot{\chi},
\ \dot{\vartheta}$. We obtain
\begin{subequations}
\label{E:4.61}
\begin{eqnarray}
q_1  &=&  L_1 / \mathcal{J}_1 \\
q_2  &=&  L_2 / \mathcal{J}_2 \\
q_3  &=&  \frac{1}{u_0}\big(L_3 -p_\varphi -2p_\chi -3 p_\vartheta
\big)  
\label{E:4.61.c} 
\end{eqnarray}
\begin{eqnarray}
\dot{\varphi} &=& \frac{p_\varphi}{2 v_0^2} 
- \frac{1}{u_0^2}\big(L_3 -p_\varphi -2p_\chi -3 p_\vartheta 
\big) \\[1pt]
\dot{\chi} &=& \frac{p_\chi}{2 u^2} 
- \frac{2}{u_0^2}\big(L_3 -p_\varphi -2p_\chi -3 p_\vartheta 
\big) \\[1pt]
\dot{\vartheta} &=& \frac{p_\vartheta}{2 \chi_0^2} 
- \frac{3}{u_0^2}\big(L_3 -p_\varphi -2p_\chi -3 p_\vartheta \big)
\end{eqnarray}
\end{subequations}
The equations~(\ref{E:4.61}) have a very simple meaning in the case 
where the potential energy does not depend on the variables $\varphi$,
$\chi$ or $\vartheta$. In such a case (a sort of model 
$\varphi$-$\chi$-$\vartheta$--instable, in the sense of the 
$\gamma$--instable model by Wilets and Jean~\cite{wilet})
the conjugate moments of these three angular variables are 
constants of the 
motion, with integer eigenvalues $n_\varphi$, $n_\chi$ and $n_\vartheta$
(in units of $\hbar$). 
Moreover, if we assume that $u_0 \rightarrow 0$,
the third component $q_3$ of the angular velocity tends to 
 $\infty$ unless $L_3 -p_\varphi -2p_\chi -3 p_\vartheta =0$
(eq.~(\ref{E:4.61.c})). 
In this case, the operator $L_3$ is diagonal, with eigenvalues
$K=n_\varphi+2n_\chi+3n_\vartheta$,
and the three degrees of freedom corresponding to
$\varphi$, $\chi$ and $\vartheta$ can be associated to non-axial
excitation modes with $K=$1, 2 and 3, respectively.

To investigate the character of the degree of freedom described by the
parameter $u_0$, and for a deeper understanding of the nature of
the other degrees of freedom, it is necessary to express the complete
Hamiltonian in the frame of a definite model which, although not
unique, is at least completely self-consistent at the limit close to
the axial symmetry.

In fact, it is now possible to use the Pauli prescriptions
\cite{pauli} to construct the 
quantum operator $\hat T$ corresponding to the classical kinetic
energy $T$ of eq.~(\ref{E:6.51c}). In doing this, we make use of the
partial inversion of the matrix $\mathcal{G}$ given by 
the solution (\ref{E:4.61}) of the linear system (\ref{E:4.7}),
and note that none of the variables $\varphi$, $\chi$, $\vartheta$ 
or $q_3$
enter in the expression of $G$:
\begin{eqnarray}
\frac{2}{\hbar^2}\ \hat{T} 
&=&-\left\{ \frac{1}{\beta_2^2 + 2 \beta_3^2} 
\frac{\partial}{\partial \beta_2}
\left[ 
\big( \beta_2^2 + 2\beta_3^2 \big) \frac{\partial}{\partial \beta_2}
\right] \right.  \nonumber \\[1pt]
&+& \frac{1}{\beta_2^2 + 2 \beta_3^2} \frac{\partial}{\partial \beta_3}
\left[ 
\big( \beta_2^2 + 2\beta_3^2 \big) \frac{\partial}{\partial \beta_3}
\right]  \nonumber \\[1pt]
&+& \frac{1}{u_0} \frac{\partial}{\partial u_0}
\left[ 
u_0 \frac{\partial}{\partial u_0}
\right] 
+ \frac{1}{2 v} \frac{\partial}{\partial v}
\left[ 
v \frac{\partial}{\partial v}
\right]  \nonumber \\[1pt]
&+& \frac{1}{2 u} \frac{\partial}{\partial u}
\left[ 
u \frac{\partial}{\partial u}
\right] 
+ \frac{1}{2 w} \frac{\partial}{\partial w}
\left[ 
w \frac{\partial}{\partial w}
\right]  \nonumber \\[1pt]
&+& \left. 
\ \frac{1}{2v^2} \frac{\partial ^2}{\partial \varphi ^2}
+ \frac{1}{2u^2} \frac{\partial 2}{\partial \chi 2}
+ \frac{1}{2w^2}
 \frac{\partial 2}{\partial \vartheta 2} \right\}  \nonumber \\[1pt]
&+& \frac{1}{4u_0^2}\ \left[ \hat{L}_3 -\frac{\partial}
{\partial \varphi}
-2 \frac{\partial}{\partial \chi} -3 \frac{\partial}{\partial
\vartheta} \right] ^2 \nonumber \\[1pt]
&+& \frac{1}{2\mathcal{J}_1}\ \hat{L}_1^2 + \frac{1}{2\mathcal{J}_2}
\ \hat{L}_2^2 \ .
\label{E:6.8}
\end{eqnarray}
This is, admittedly, only a semi-classical discussion. However, the
formal quantum treatment in the frame of the Pauli procedure, 
shown in the Appendix \ref{S:a4}, gives exactly the same result.

Until now, no assumption has been made on the form of the
potential-energy operator, which will determine the particular model. 
A few general remarks on this subject are contained in the Appendix
\ref{S:a3}. We now assume that the potential energy can be separated
in the sum of a term depending only on $u_0$ and another
containing the other dynamical variables. In this case the
differential equation in the variable $u_0$ is approximately
decoupled from the rest. One obtains the Schr\"odinger equation
\begin{eqnarray}
\left\{ \frac{1}{u_0} \frac{\partial}{\partial u_0}
 \left[u_0 \frac{\partial}{\partial u_0} \right] \right. &+&
\frac{2}{\hbar^2}
\left[ E_{u_0} - U(u_0) \right]  \nonumber \\
-\ \frac{1}{u_0^2}&& \left. \left[ 
\frac{\Omega_{u_0}}{2}\right] ^2
 \right\} \phi (u_0) =0
\label{E:6.9}
\end{eqnarray}
where we have put
\begin{equation}
\Omega_{u_0} = K - n_\varphi -2n_\chi -3n_\vartheta \ .
\label{E:6.10}
\end{equation}
If we assume, for simplicity, a harmonic form for the potential
$U(u_0)=\frac{1}{2}C u_0^2$, the eq.~(\ref{E:6.9}) is the
radial equation of a bidimensional harmonic oscillator. For the
existence of a solution, it is required that
\begin{eqnarray}
\Omega_{u_0}&=& 2n_{u_0} \nonumber \\ 
E_{u_0}&=& (N_{u_0}+1) \hbar \omega_{u_0}
\end{eqnarray} 
with $n_{u_0}$ positive  or negative integer and 
the integer $N_{u_0}\geq |n_{u_0}|$.
Excitations in the degree of freedom corresponding to the variable
$u_0$ carry, therefore, two units of angular momentum in the
direction of the 3rd axis of the intrinsic reference frame.

We could extend our model to include all the intrinsic variables
different from $\beta_2,\ \beta_3$. We assume a potential energy
corresponding to the sum  of independent harmonic potentials in the
variables $v ,\ u,\ w$ and $u_0$, plus a term depending
on $\beta_2$ and $\beta_3$ (at the moment, we do not need to define the
form of this term). We also assume that the equations in 
$\beta_2,\ \beta_3$ can be approximately decoupled from those of the
other variables and that, in the latter, $\beta_2$ and $\beta_3$
can be replaced by their average values. It is easy to verify that the
differential equations in the pair of variables $v ,\ \varphi$
(or $u,\ \chi$ or $w,\ \vartheta$) correspond again to a
bidimensional harmonic oscillator and that, as long as we
neglect the rotation--vibration coupling, the eigenvalue
$K$ of the intrinsic component $L_3$ of the angular momentum is given
by
\begin{equation}
K= n_\varphi + 2n_\chi +3n_\vartheta +2n_{u_0}\ .
\end{equation}
The energy eigenvalues are, for the equation in the variables 
$v ,\ \varphi$,
\begin{equation}
E_v=(N_v +1)\ \hbar \omega_v \hspace{8mm}{\rm with}
\ N_v \geq |n_\varphi |
\end{equation}
and have a similar form for the other two oscillators.

It remains to consider the character of the different dynamical
variables with respect to the parity operator.
We know that the parity of the amplitude $a^{(\lambda )}_\mu$ is
$(-1)^\lambda$. Therefore, $\beta_2$ and $\gamma_2$ are even, while
$\beta_3$, $X$ and $Y$ are odd. As for $\gamma_3$, we observe that 
$\beta_3 \sin \gamma_3$ is odd, and therefore  $\gamma_3$ must be
even.
As a consequence, are also even the linear combinations $\gamma$ and
$\gamma_0/c$ defined in the eq.~(\ref{E:4.1a},\ref{E:4.2}).
The new variables $\eta ,\ \zeta$ and $\xi$ are odd, as (\textit{e.g.})
$\beta_2 \xi $ is an octupole amplitude and therefore is odd, while
$\beta_3 \xi$ must be even. Finally, on the basis of eq.s~(\ref{E:4.5})
we realize that $v ,\ u$ and $w$ must be odd 
(while $\varphi ,\chi$ and $\vartheta$ are even). We have therefore
identified elementary excitations corresponding to 
$K^\pi =1^-,2^-,3^-$ and $2^+$. Excitations with $K=0$ (of positive or
negative parity) conserve the axial symmetry, and are related to the
variables $\beta_2$ and $\beta_3$. A particular example will be
discussed in the following Sections. 

\section{\label{S:3} A specific model: 
Axial octupole vibrations in nuclei with
permanent quadrupole deformation}

Specific assumptions on the form of the
potential--energy terms for all the variables describing the
quadrupole and octupole degrees of freedom are necessary in order to
obtain definite predictions, also if these are limited to the axial
modes.

We discuss here, as an example, the case of axial octupole excitations
in nuclei which already possess a stable quadrupole deformation. In
this case, one obtains relatively simple results, suitable for
comparison with experimental data. This comparison will be performed
in the next Section.

Following the usual treatment of vibration + rotation, we
put $\beta_2 = \bar{\beta}_2+\beta_2^\prime$, with $\bar{\beta}_2=$
constant and $|\beta_2^\prime |\ll |\bar{\beta}_2|$.
The new variable $\beta_2^\prime$ is therefore assumed to be small
(of the first order) as all other variables, with the exception of
$\beta_3$.
With this choice, and assuming that the variables introduced in the
eq.~(\ref{E:4.5}) are suitable to describe the other degrees of freedom,
the matrix $\mathcal{G}$ takes a form similar to that of Table II (with 
$\beta_2$ substituted by $\bar{\beta}_2$ and $\dot{\beta}_2$ by 
$\dot{\beta}_2^\prime$) and -- at the lowest significant order --
results to be diagonal with respect to the variables $\dot{\beta}_3$,
$\dot{\beta}_2^\prime$, $q_1$ and $q_2$.
Moreover, in our model, the amplitude of oscillation for all
degrees of freedom different from $\beta_3$ are constrained to very
small values: this fact implies strong restoring forces and,
therefore, oscillation frequencies much larger than for $\beta_3$. 

In the limit of small amplitude of the octupole oscillations, this
case has been discussed, \textit{ e.g.}, by Eisenberg and
Greiner~\cite{eisen1}.
In their approach, the ``intrinsic'' reference frame is chosen to
coincide with the principal axis of the quadrupole deformation tensor,
but the differences between their approach and ours tend to disappear
for $|\beta_3| \ll \bar{\beta}_2$. The model we try to develop should
be able to describe (axial) octupole vibrations of finite amplitude,
but its limit for  $|\beta_3| \ll \bar{\beta}_2$ must obviously agree
with the results of Eisenberg and Greiner.

The quantum-mechanical equation of motion for $\beta_3$ can be
obtained with the Pauli prescription~\cite{pauli}, with the additional
assumption that the equations involving $\beta_3$
and $\beta_2^\prime$ or the angular-momentum components $L_1,\ L_2$ are
effectively decoupled from those containing the other dynamical
variables and/or the $L_3$ operator. The latter equations could
possibly be complicated, and substantially coupled with one another
and with the angular momentum component $L_3$ along the (approximate)
symmetry axis. A short discussion of this subject, with some
simplifying assumptions, has been given in the previous
section~\ref{S:2.4}.
At the moment, we assume  that terms involving $\beta_2^\prime,\ L_3$
and other dynamical variables different from $\beta_3$ contribute to
the total energy with their own eigenvalue, 
 independent from the
eigenfunction in the $\beta_3$ degree of freedom, and we only consider
their lowest--energy state. We also assume that this state has $K=0$
(and neglect, as  usual, the possible rotation-vibration coupling). In
this case the complete wavefunction has the form
\begin{eqnarray}
\Psi &\propto& \psi (\beta_3) \Phi_0(\beta_2^\prime,\gamma_0,w,\rho ,
\xi_0 ,\vartheta ,\chi ,\varphi )
\mathcal{ D}^J_{M,0}\nonumber \\*
& \propto& \psi (\beta_3) \Phi_0\ Y_{JM}(\hat{\Omega})
\label{E:5.1}
\end{eqnarray}
where $\mathcal{ D}^J_{M,M^\prime}$ are the Wigner matrices and
 $\psi (-\beta_3)=(-)^J \psi (\beta_3)$. The differential equation
for the wavefunction $\psi (\beta_3)$, obtained with the Pauli
prescription, has the form
\begin{eqnarray}
\left[ -\frac{\hbar^2}{2}\right.&G^{-1/2}&\frac{\rm d}{{\rm d}\beta_3}
G^{1/2}\frac{\rm d}{{\rm d}\beta_3} + V(\beta_3)  \nonumber \\*
 &+&\left. \!\frac{\hbar^2}{2}
\frac{J(J+1)}{\mathcal{ J}_1} \right] \psi (\beta_3)=E \psi (\beta_3)
\label{E:5.2}
\end{eqnarray}
where $V(\beta_3)$ is the potential-energy term.
The expression of the determinant $G$ is not uniquely defined, as it
also depends on the part of the matrix of inertia $\mathcal{G}$ 
involving all
other dynamical variables. This is a general problem for all models
where part of the degrees of freedom are ignored (and, in collective
models, a number of dynamical variables describing the details of
nucleon degrees of freedom, are certainly ignored).
With the present choice of dynamical variables (Table~\ref{T:2}), 
one gets
$G=$Det$\;\mathcal{ G}=2304\ u_0^2\ v^2 u^2 w^2 
(\bar{\beta}_2^2 + 2
\beta_3^2)^2$,
and the eq.~(\ref{E:5.2}) becomes
\begin{eqnarray}
\left[ -\frac{\hbar^2}{2}\! \right.&&\!\left( \frac{{\rm d}^2}
{{\rm d}\beta_3^2} +
\frac{4\beta_3}{\bar{\beta}_2^2+2\beta_3^2}\frac{\rm d}{{\rm d}\beta_3}
\right) + V(\beta_3)\\
 &+& \left. \frac{\hbar^2}{2}
\frac{J(J+1)}{\mathcal{ J}_1} \right] \psi (\beta_3)=E \psi (\beta_3)\
.
\label{E:5.3}
\end{eqnarray}
This is the equation we have used in ref.s~\cite{bizoct1,bizoct2}.
A different choice of the dynamical variables would have brought to a
different result. \textit{ E.g.}, with the variables used in the
Table~\ref{T:1},
one obtains\\ 
$G\propto \bar{\beta}_2^2 \beta_3^2 (\bar{\beta}_2^2 +2\beta_3^2)^2
(\bar{\beta}_2^2 +5\beta_3^2)^{-1} (\bar{\beta}_2^2 \gamma_2+
\sqrt{5}\beta_3^2
\gamma_3)^2$.\\
However, in this case the limit $|\beta_3| \ll |\bar{\beta}_2|$
would not correspond to the Eisenberg--Greiner result, due to the
presence of the factor $\beta_3^2$ in the expression of the
determinant $G$. In fact, when the spherical symmetry is broken by the
permanent quadrupole deformation, and at the limit of small octupole
deformation, the octupole amplitudes $a^{(3)}_\mu$ are decoupled from
one another~\cite{eisen1} and it would have been more reasonable 
to chose a
dynamical variable $u_3=\beta_3 \gamma_3$ in the place of $\gamma_3$.
With this substitution, the factor $\beta_3^2$ disappears.

It is convenient to express the differential equation~(\ref{E:5.3}) in
terms of the adimensional quantities
$x=\sqrt{2}\beta_3/\bar{\beta}_2$ and $\epsilon =
\frac{1}{\hbar^2}\bar{\beta}_2^2 E$,
$v=\frac{1}{\hbar^2}\bar{\beta}_2^2 V$. One obtains
\begin{equation}
\frac{{\rm d}^2\psi (x)}{{\rm d}x^2} +
\frac{2x}{1+x^2}\frac{\rm d\psi (x)}{{\rm d}x}
 + \left[ \epsilon -
\frac{J(J+1)}{6(1+x^2)} -v(x)\right] \psi (x)\!=\!0
\label{E:5.4}
\end{equation}
where $\psi (-x)=(-)^J \psi (x)$.
This equation reduces to that of Eisenberg and Greiner~\cite{eisen1}
when $x \ll 1$.

As for the potential $v(x)$, we have explored two possible
forms: a quadratic term\footnote{After completion of this work, we
have been informed that a quadratic potential plus a centrifugal
term with variable moment of inertia has also been considered in the
model by Minkov \textit{et al.}~\protect{\cite{minkov04}}.}
 $v(x)=\frac{1}{2}c x^2$ or a
square--well potential, as it has been adopted~\cite{iac2} 
at the critical point in the X(5) model ( $v(x)=0$ for
$|x|<b$ and $=+\infty$ outside, so that $\psi(\pm b)=0$ ).
In both cases, there is a free parameter to be determined from the
comparison with empirical data.

We now discuss in particular the second case.
We may note that for $V=0$ the eq.~(\ref{E:5.4}) is formally
equivalent to that of \textsl{ spheroidal oblate wavefunctions} (see
eq.~21.6.3 of ref.~\cite{abram}) with the parameters $m,\ \lambda $ and
$C$ redefined as $m=0$, $C^2=\epsilon$ and
$\lambda =J(J+1)/6 -\epsilon$. Here, however, the solution is confined
in the interval $-b < x < b$ and the equation has been  solved
numerically.
\begin{figure}[th]
\resizebox{8.4cm}{!}
{\includegraphics[80,468][556,743]{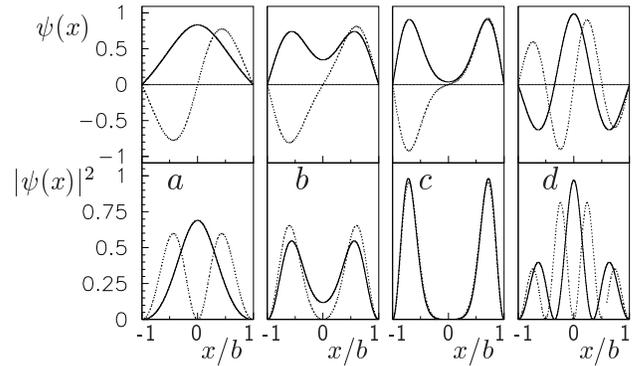}}
\caption{\label{F:1} Examples of wavefunctions $\psi_{sJ}(x)$  
as a function of
$x/b$, for $b=1.73$. Pairs of curves refer to consecutive values of
$J$, with even parity  (continuous line) or  odd parity (dotted line).
Part $a:\ J^\pi=0^+,1^-$,
 $b:\ J^\pi=10^+,11^-$, $c:\ J^\pi=18^+,19^-$ with $s=1$;
$d:\ J^\pi=0^+,1^-$ with $s=2$.
With increasing angular momentum, the difference in $|\psi(x)|^2$ between
consecutive values of $J$ tends to vanish and, as a consequence, the
positive and negative part of the band merge together.}
\end{figure}
For a given $b$ and for every value of $J$, one obtains a complete set
of orthogonal eigenfunctions, with an integrating factor $(1+x^2)$.
These eigenfunctions can be characterized by the quantum
number $s=\nu +1$, where $\nu$ is the number of zeroes in the open
interval $0 < x <b$. A few examples of wavefunctions corresponding to
the square--well potential with $b=1.73$ and for different values of
$s$ and $J$ are depicted in the Fig.~\ref{F:1}.

\begin{figure}[t]
\resizebox{7.5cm}{!}
{\includegraphics[35,395][540,707]{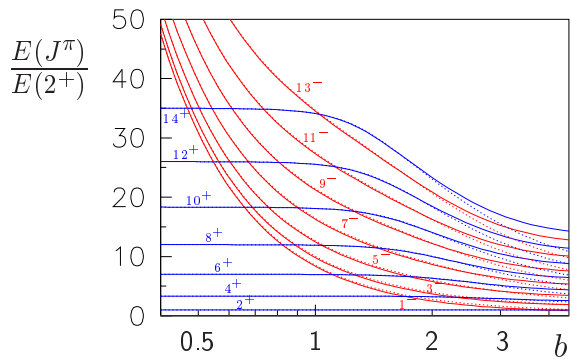}} 
\caption{\label{F:2} (Color online) Ratio   
$E(J^\pi)/E(2^+)$ as a function of $b$, for
states of the ground--state ($s=1$)  band
 with different $J^\pi$. Dotted lines 
show the results obtained with a differential equation corresponding
to that of Eq.~(\ref{E:5.4}) \textsl{ but 
without the first--derivative term}.}
\end{figure}

The dependence of the eigenvalues on the parameter $b$ is illustrated
by the Fig.~\ref{F:2}, where the ratios $E(J)/E(2)$ are shown for the
g.s. band ($s=1$). 
 Other possible choices of the set of independent
dynamical variables would have brought to a different equation, but
the difference would have concerned the coefficient of the
first--derivative term the eq.~(\ref{E:5.4}),
 with very small effect on the results, as long
as $b$ is in a range of ``reasonable'' values.
To exemplify the effect of this term, results obtained with the
coefficient of the first derivative put to zero are also shown -- as
dotted lines -- in the Figure~\ref{F:2}. Differences between the two
sets of
results turn out to be very limited for small values of $b$ (at least,
up to $b\approx 2$).
There is at least one case ($^{226}$Th) in which our results
with the square--well 
potential are in good agreement with the level scheme, for low-lying
states of positive and negative parity, while for $^{228}$Th 
a better agreement is obtained with the quadratic potential. 
The possible interpretation of this result as evidence for a phase
transition in the octupole degree of freedom is discussed in the
next Section.

\section{\label{S:4} Evidence of phase transition 
in the octupole degree of freedom}
\subsection{\label{S:4.1} The Radium and Thorium isotopic chain}

A phase transition in the nuclear shape manifests itself as a
relatively sharp change of a proper {\em order parameter} --\textit{
e.g.}, the ratio $R=E(4^+)/E(2^+)$ -- as a function of a 
\textsl{ driving
parameter} which can be, in our case, the number of neutrons in the
isotopes of a given element or the number of protons along an isotone
chain. Due to the finite number of degrees of freedom, the transition
region has a finite width around the critical point, and extends over
several nuclides in the chain. In the case of transitions between
spherical shape and axial quadrupole deformation, the X(5) symmetry,
 valid at the critical point, predicts
$E(4^+)/E(2^+)=2.91$ and we can use this criterion to locate the
critical point of the phase transition.
\begin{figure}[h]
\rotatebox{90}
{\resizebox{!}{8.5cm}
{\includegraphics[75,28][518,750]{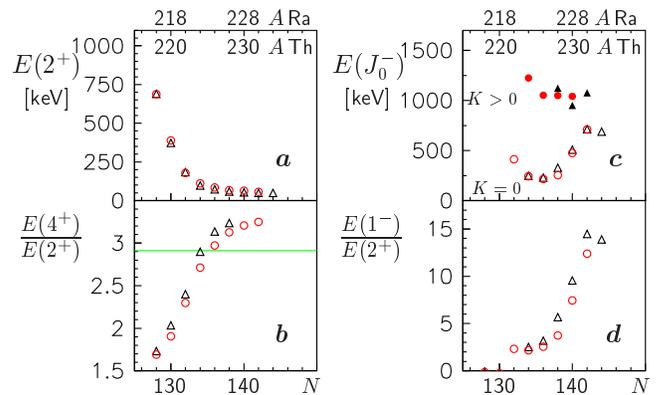}}}
\caption{\label{F:3} (Color online) Indicators of the quadrupole 
collectivity (left)
and of the octupole collectivity (right), as a function of the neutron
number $N$ in the isotopic chain of Ra (circles) and Th (triangles):
$a$ - Excitation energy of the first $2^+$ level; $b$ - Energy ratio
$E(4^+)/E(2^+)$; $c$ - Excitation energy of the first level of the
$K^\pi=0^-$ band, $J_0^\pi=1^-$ (open symbols) and of the lowest known
level of other negative-parity bands, $J_0^\pi=2^-$ or $1^-_2$ (full
symbols); $d$ - Energy ratio $E(1^-)/E(2^+)$.
The horizontal line in the part $b$ shows the value (2.91) 
expected for the X(5) symmetry. Data are from ref.s~\cite{cocks,nndc}. }
\end{figure}
The situation is more complex when the quadrupole and the octupole
degrees of freedom must be considered at the same time. Fig.~\ref{F:3}
shows, as a function of the neutron number of Ra and Th isotopes, a
few parameters which can be used as indicators of the quadrupole and
octupole collectivity. As for the quadrupole mode, the decrease of
$E(2^+)$ with increasing $N$ (Fig.~\ref{F:3}$a$) shows a corresponding
increase of collectivity.
Moreover, in the Fig.~\ref{F:3}$b$ 
we observe the transition between the vibrational (or not
collective) behavior of the lighter isotopes of the chain 
and a clear rotational
behavior ($R \approx 10/3$) above $A=226$, with a critical point
 which can be located around $A=224$.
The ratio $E(1^-)/E(2^+)$, depicted in Fig.~\ref{F:3}$d$, shows that the
relative importance of the octupole collectivity increases with
decreasing $N$ and reaches its maximum in the region below $N=138$,
where the
critical point of the phase transition in the quadrupole mode could be
located on the basis of Fig.~\ref{F:3}$b$.
Heavier isotopes show evidence of octupole vibrations 
(of different $K$) around a quadrupole-deformed core~\cite{guent00,raduta}.
Lighter isotopes ($N<132$)
appear not to be deformed in their lower-$J$ states.
However, at larger angular momentum a rotational-like band develops,
and this band has the alternate-parity pattern typical of a
stable octupole deformation~\cite{smith}.

The model introduced in the first part of this paper, and developed in
the section~\ref{S:3} for the particular case of a permanent quadrupole
deformation, assumes that non--axial amplitudes are constrained to
very small values by the large restoring forces. This implies that
excitation of one of the non--axial degrees of freedom leads to high
excitation energy, compared to that of 
the first $1^-$ level of the $K=0$ band. Experimental data of 
Fig.~\ref{F:3}$c$
show that this is actually the case for the light Thorium isotopes, at
least up to $A=228$. In fact, the first $1^-$ level is not far  from
the first
$2^+$ and much lower than levels belonging to negative--parity bands 
with $K\neq 0$ (as the lowest $2^-$ or the second $1^-$).

\begin{figure*}[t]
\resizebox{15cm}{!}
{\includegraphics[0,0][480,204]{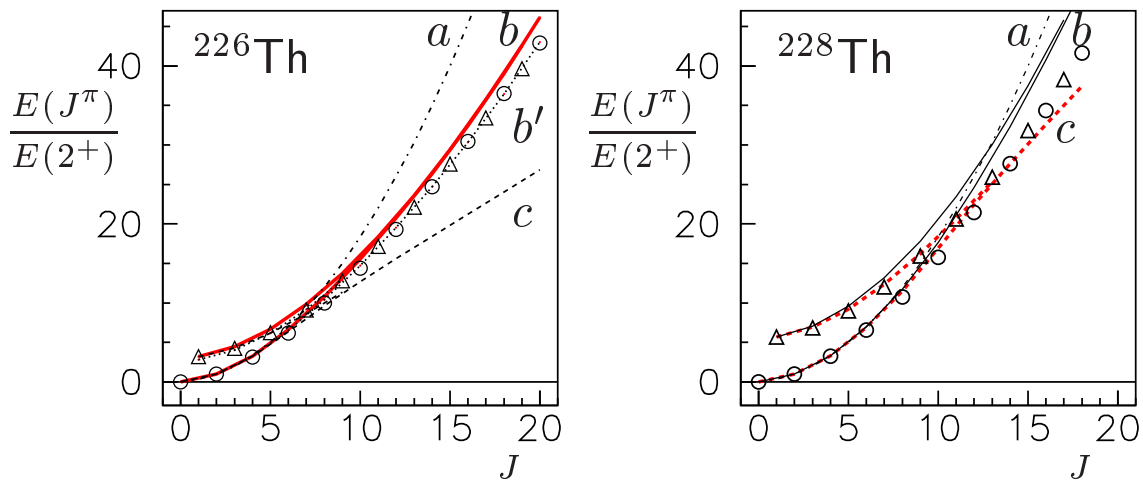}}
\caption{\label{F:4} (Color online) Ratios   
$E(J^\pi)/E(2^+)$ 
as a function of $J$,
for positive-parity states (circles) and 
negative-parity states (triangles)
of the ground--state ($s=1$) band of $^{226}$Th and $^{228}$Th,
compared with different model calculations: rigid rotor (curve $a$,
even parity only),
present model at the critical point (curves $b$ and $b^\prime$), 
present model with harmonic potential in $\beta _3$ (curve $c$).
The curves $b$ and $c$ correspond to a fit on the lowest $1^-$ state,
the curve $b^\prime$, on the $20^+$ state.
Note that the even-parity parts of the curves $b$ and $b^\prime$
are very close to the curve $a$ for $J < 6$, while,
for each one of the curves $b$, $b^\prime$ and $c$, 
the even- and odd-parity branches  tend to merge together 
at large values of J.
}
\end{figure*}
\subsection{\label{S.4.2} Comparison with experimental 
data for $^{226,228}$Th}
Our model assumes a permanent quadrupole deformation. Therefore, it
can be useful only for relatively heavy Th isotopes 
(Fig.~\ref{F:3}$a,b$). The quadrupole-deformed
region extends above the mass 224 (which could correspond to the 
critical point of the phase
transition, having $E(4^+)/E(2^+)\approx 2.91$). Heavier Th isotopes
(with $A\geq 230$) show negative-parity bands built on the different 
states of octupole vibration, from $K=0$ to $K=3$, with band heads much
higher than the first $2^+$ (Fig.~\ref{F:3}$c$). 
Only for lower $A$, the $1^-$ band head
of the $K^\pi=0^-$ band decreases well below the band heads of all
other octupole bands, and higher levels of the  $K^\pi=0^-$ band
merge with those
of positive-parity of the ground-state band (Fig.~\ref{F:4}), 
approaching (but not
reaching) the pattern expected for a rigid, reflection asymmetric
rotor.

The region of possible validity of our model is therefore restricted 
to $^{226}$Th and $^{228}$Th.

The ratios $E(J^\pi)/E(2^+)$ for the low-lying states of   
$^{226}$Th and $^{228}$Th are depicted in the Fig.~\ref{F:4}, and
compared with the predictions of different models. The rigid-rotor
model cannot account for the position of the lowest negative-parity
levels, and overestimates the excitation energy for all the high-spin
states. Instead, a rather good agreement is obtained with the present
model, if one assumes, in the case of $^{226}$Th, a square-well
potential (as the one hypothesized by Iachello in his X(5) model)
and, in the case of $^{228}$Th, a harmonic restoring force.
In both cases, the free parameter of the model has been adjusted to
reproduce the position of the $1^-$ level. As shown in the Fig.
\ref{F:4} (dotted line) a much better agreement with the high-spin
levels of $^{226}$Th is obtained with a slightly different value of
the parameter ($b=1.87$ instead of 1.73), 
at the expense of a very limited discrepancy
for  the $1^-$ level.

Therefore, for what concerns the level energies of the ground-state
band (including in it also the odd-$J$, negative-parity states),
 $^{226}$Th seems to present the expected behavior of a nucleus with
permanent quadrupole deformation and close to the critical point of the
phase transition in the octupole mode.

\subsection{\label{S.4.2a} 
Other possible tests of the critical-point behavior}

A considerable amount of experimental information has been reported,
in the last few years, on possible 
candidates~\cite{casten1,krue1,biz1,hutt,clar,fran,bijk03,bijk04,tone}
 for the dynamical
symmetry X(5)(phase transition point in the quadrupole mode). 
It is now clear that the agreement between experimental and calculated
energies for the ground-state band does not automatically imply that a
similar agreement exists also for other observables, like the excitation
energy of the second $0^+$ level (the band head of the $s=2$ band)
and the in-band and inter-band transition probabilities. 
In several transitional nuclei, the excitation energies in the
ground-state band are in excellent agreement with the X(5)
predictions but the calculated ratios of the $B$(E2) transition
probabilities fail to reproduce the experimental ones~\cite{clar},
unless an \textit{ad-hoc} second-order term is included in the E2
transition operator~\cite{arias01,capr04}.
It is therefore important to test the predictions of our models
also for what concerns such observables.

\begin{figure}[t]
\resizebox{8cm}{!}
{\includegraphics[20,385][412,705]{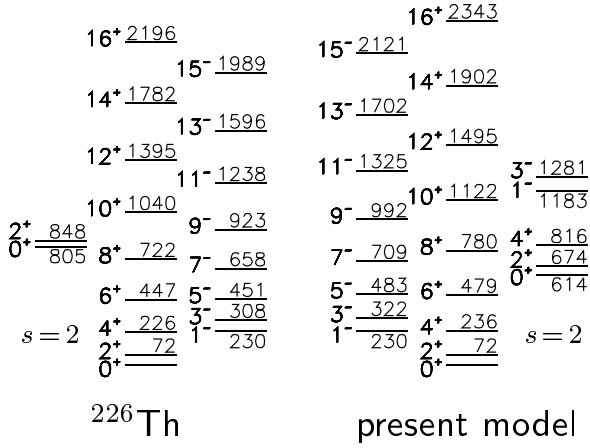}}
\caption{\label{F:scheme}
Experimental level scheme of $^{226}$Th compared with the 
predictions of the present model at the critical point. The
two spectra are normalized on the first $2^+$ level, and the
model parameter $b$ is adjusted to reproduce the position of 
the first~$1^-$.} 
\end{figure}

The low--lying level scheme of $^{226}$Th is shown in the 
Fig.~\ref{F:scheme}, together with the one resulting from the present
model, with the value of $b$ adjusted in order to reproduce the 
empirical value of the ratio $E(1^-)/E(2^+)$.
At the moment, only the first two levels of the $s=2$ band 
($0^+_2,\ 2^+_2$) 
are known and their excitation energies are somewhat higher than the
values predicted at the critical point.
We can observe, however, that also in the best X(5) nuclei
\cite{casten1,krue1,tone} the
position of the levels of the $s=2$ band deviates somewhat from the
model predictions (although in the opposite direction). 
In our opinion, a similar
qualitative agreement is obtained also in the present case. The
negative--parity levels of the $s=2$ band are predicted to lie at 
higher energies, and could be difficult to observe.
Absolute values of the transition strengths are not available for
$^{226}$Th, but some relevant information is provided by the branching
ratios in the level decays. In the tables \ref{T:e1},\ref{T:e2}, 
experimental ratios of the reduced transition strengths  for E1 or E2
transitions coming from the same level are compared with the model
prediction at the critical point.
{\renewcommand{\arraystretch}{1.10}
\begin{table}[h]
\caption{\label{T:e1} Experimental and calculated values 
of the ratios of 
reduced strengths for E1 transitions
coming from the same level of $^{226}$Th.}
\begin{ruledtabular}
\begin{tabular}{lrlrlccrl}
&\multicolumn{2}{l}{Trans. 1}&\multicolumn{2}{l}{Trans. 2}
&\multicolumn{4}{c}{$ B_{i\rightarrow f_1}$
(E1)/$B_{i\rightarrow f_2}$(E1) }\cr 
$J^\pi_i$ & \multicolumn{2}{c}{$\Rightarrow J^\pi_{f_1}$}
& \multicolumn{2}{c}{$\Rightarrow J^\pi_{f_2}$} &
\multicolumn{1}{l}{Theoretical} & \multicolumn{3}{c}{Experimental} 
\\[1pt]
\hline
& & & & & & & \\[-8pt]         
$1^-$ & & $\Rightarrow 0^+$ & & $\Rightarrow 2^+$ 
& {0.47} & &{0.54} & {(5)}\cr
$3^-$ & & $\Rightarrow 2^+$ & & $\Rightarrow 4^+$ 
& {0.65} & &{0.99} & {(25)}\cr
{$2_2^+$} & & $\Rightarrow 1^-$ & & $\Rightarrow 3^-$ 
& {0.63} & & {0.60} & {(18)}\cr
\end{tabular}
\end{ruledtabular}
\end{table}
}
\smallskip 

\begin{table}[h]
\caption{\label{T:e2} Experimental and calculated values of the ratios
of  reduced strengths (in W.u.) $R=B(E1)/B(E2)$, for transitions
coming from the same level of $^{226}$Th.}
\begin{ruledtabular}
\begin{tabular}{lrlrlccrl}
&\multicolumn{2}{l}{Trans. 1}&\multicolumn{2}{l}{Trans. 2}
&\multicolumn{4}{c}{$R\ \times \ 10^5$}\\ 
$J^\pi_i$ & \multicolumn{2}{c}{$\Rightarrow J^\pi_{f1}$}
& \multicolumn{2}{c}{$\Rightarrow J^\pi_{f2}$} &
\multicolumn{1}{l}{Theoretical} & \multicolumn{3}{c}{Experimental} 
\\[2pt]
\hline         
$8^+$ & E1 & $7^-$ & E2 & $6^+$ &  1.3 & & 2.0 & (8)\\
$9^-$ & E1 & $8^+$ & E2 & $7^-$  & 1.3 & & 1.7 & (2)\\
$10^+$ & E1 & $9^-$ & E2 & $8^+$ & 1.6 & & 1.5 
& (1)\footnotemark[1]\\
$11^-$ & E1 & $10^+$ & E2 & $9^-$ &  1.6 & & 1.7 
& (1)\footnotemark[1]\\
$12^+$ & E1 & $11^-$ & E2 & $10^+$ &  1.8 & & 1.6 & (1)\\
$13^-$ & E1 & $12^+$ & E2 & $11^-$ &  1.8 & & & \\
$14^+$ & E1 & $13^-$ & E2 & $12^+$ &  1.9 & & 1.4 & (1)\\
$15^-$ & E1 & $14^+$ & E2 & $13^-$ &  2.0 & & 1.7 & (3) \\
$16^+$ & E1 & $15^-$ & E2 & $14^+$ &  2.1 & & & \\
$17^-$ & E1 & $16^+$ & E2 & $15^-$ &  2.1 & & 1.5 & (3) \\
$18^+$ & E1 & $17^-$ & E2 & $16^+$ &  2.2 & & & \\
$19^-$ & E1 & $18^+$ & E2 & $17^-$ &  2.3 & & 1.7 & (4)\\
\end{tabular}
\end{ruledtabular}
\footnotetext[1]{ Values used for normalization.}
\end{table}
The electric dipole moment would vanish for the collective motion of
a fluid with uniform charge density, as the center of charge would
coincide with the center of mass. Therefore,the observed E1 transition
amplitudes are entirely due to the non uniformity of the 
nuclear charge distribution~\cite{tzvet02}.
To calculate the value of 
$B$(E1)$=(i|| \mathcal{M}(E1)||f)^2 /(2J_i+1)$, 
the  E1 transition operator has been assumed to have the 
form~\cite{bohr57,bohr58,stru57,lipas63}
\begin{equation}
\mathcal{ M}_\mu (E1) = C_1 \beta_2 \beta_3 Y_{1,\mu}\ ,
\label{E:3.1}
\end{equation}
with the constant factor  $C_1$ depending on the nuclear charge
polarizability. 
The E2 transition operator for the in-band transition and at the limit
close to the axial symmetry has been taken in the simple form
\begin{equation}
 \mathcal{M}_\mu (E2) = C_2(\beta_0 ) Y_{2,\mu}\ ,
\end{equation}
neglecting the (weak) dependence on $\beta_3^2$.
Therefore, the theoretical ratio of the reduced strengths for
transitions of different multipolarity (E1 and E2) is determined apart
from a constant factor, which must be fixed by comparison with the
experimental data. The average of the ratios
$B$(E1)/$B$(E2) for the transitions coming from the $10^+$ and $11^-$
levels has been used for normalization of the theoretical values given
in the Table \ref{T:e2}.

\section{\label{S:5} Conclusions}
A theoretical scheme for the description of quadrupole plus octupole
excitations close to the axial symmetry limit has been developed in
the Section~\ref{S:2} and specialized in the Section \ref{S:3} to the
simpler case of a permanent (and axially symmetric) quadrupole
deformation. In principle, the model should be able to describe the
wide field  of reflection--asymmetric nuclear shapes, close to (but
not coincident with!) the axial symmetry limit. Calculations of
nuclear shapes in the frame of the HBF-Cranking model~\cite{naz} for
nuclei of the Radium -- Thorium region find a large variety of
results, including proper potentials
for quadrupole--octupole vibrations around a spherical
shape or for octupole vibrations around a deformed, reflection symmetric
shape, and also situations with a rather flat minimum of the potential
along a line at constant $\beta_2$  with $|\beta_3 | < \beta_3^{max}$.
The latter case
is just what is expected for our ``critical point'' of the 
phase transition in the octupole mode. 
In the Section \ref{S:4}, we have investigated the
evolution of the nuclear shape along the isotopic chain of Thorium,
and shown that evidence of phase transition exists, not only in the
quadrupole mode but also in the octupole mode around a stable
quadrupole deformation. The model developed in the Section \ref{S:3}
results to be able to account for the experimental data of $^{226}$Th
(at the critical point of the phase transition in the 
octupole mode), and also of its
neighbor  $^{228}$Th (characterized by axial octupole
vibrations). More and improved experimental data on E2 and E1
transition strengths would be necessary for a more stringent test of
the model predictions.

Further developments of the calculations are in program, to provide
detailed predictions for other significant cases: first of all, those
of  $^{224}$Ra and  $^{224}$Th, whose positive--parity levels show an
energy sequence very close to Iachello X(5) predictions 
 for the critical point of phase transition between 
spherical shape and axially symmetric quadrupole deformation.

\appendix

\section{\label{S:a1} Summary of the general formalism}

The general formalism to describe collective states of
rotation/vibration in nuclei is discussed \textit{ e.g.} in
ref.~\cite{bohr52}.
The nuclear surface is described in polar coordinates as 
\begin{equation}
r(\theta ,\phi ) = R_0 \left[ 1 + \sum _\lambda 
\sum _{\mu = -\lambda, \lambda} 
\alpha ^{(\lambda )}_\mu Y^*_{\lambda, \mu}(\theta ,\phi ) 
\right] ,
\label{E:a1}
\end{equation}
with the condition $\alpha ^{(\lambda )}_{-\mu} =
(-1)^\mu {\alpha^{(\lambda )}_\mu}^*$.

In the sum, the values of $\lambda$ are now limited to 2,3.
A term with $\lambda = 1$ should be included in order to 
maintain fixed
the position of the center of mass~\cite{eisen1}:
\begin{eqnarray}
\alpha^{(1)}_\mu &=& \frac{3}{\sqrt{16 \pi}}\ 
\sum _{\lambda ,\lambda^\prime}
\sqrt{(2\lambda +1) (2\lambda^\prime +1)} \nonumber \\
&&\cdot \left( \begin{array}{ccc}
\lambda & \lambda^\prime & 1 \\
0 & 0 & 0
\end{array} \right)
\ \left[ \alpha^{(\lambda )} \otimes  \alpha^{(\lambda^\prime )}
\right] ^{(1)}_\mu \ .
\end{eqnarray}
This term, however, is inessential
for the following discussion, and will be omitted.
For not-too-large deformation, in fact, the amplitudes 
$\alpha^{(1)}_\mu$ are much smaller than the others, and their effect 
results to be negligible for most purposes (with the noticeable 
exception of the E1 transition amplitudes). At the same level of
accuracy,
one can also neglect the slight variation of $R_0$ necessary to keep
the volume exactly constant.

In the Bohr model that we are considering here, the classical
expression of the collective kinetic energy is
\begin{equation}
T = \frac{1}{2} B_2 \sum _{\mu = -2,2} \left| \dot{\alpha}^{(2)}_\mu 
\right| ^2
+ \frac{1}{2} B_3 \sum _{\mu = -3,3} \left| \dot{\alpha}^{(3)}_\mu 
\right| ^2 \ .
\label{E:a2}
\end{equation}
In order to simplify the notation in the following, we include the 
inertia coefficient $B_\lambda$ in the definition of the collective 
variables $a^{(\lambda)}_\mu $.
In the literature, this symbol is usually reserved
to the variables defined in the intrinsic reference frame and, from
now on, we will always use this reference frame for the collective
variables $a^{(\lambda)}_\mu$:
\begin{equation}
\sqrt{B_\lambda}\ \alpha^{(\lambda)}_\mu =
\sum_\nu a^{(\lambda)}_\nu {D^{(\lambda )}_{\mu \nu}}^*(\theta_i ) 
\label{E:a.3}
\end{equation}
where $D^{(\lambda )}$ are the Wigner matrices and $\theta_i$ the Euler angles.
The axis of the intrinsic reference frame are defined along the
principal axis of the inertia tensor. 

The expression of the kinetic energy
in terms of the time derivatives
of the intrinsic deformation variables  $ a^{(\lambda)}_\mu $
and of the angular velocity $\vec q$ of the intrinsic frame with
respect to an inertial frame is discussed, \textit{e.g.}, in the ref.s
\cite{bohr52,eisen1}.
If only the quadrupole mode is considered, the total kinetic energy
can be expressed as the sum of a vibrational term (in the intrinsic
frame) and a rotational term. If the octupole mode is also considered,
a rotation-vibration coupling term must be added~\cite{eisen1,wexler}:
\begin{equation}
T = T_{\rm vib} + T_{\rm rot} + T_{\rm coup}
\label{E:a.5}
\end{equation}
where
\begin{eqnarray}
T_{\rm vib} &=& \frac{1}{2}
\sum _{\lambda ,\mu} \left| \dot{a}^{(\lambda )}_\mu \right| ^2\\
T_{\rm rot} &=& \frac{1}{2}
\sum_{\lambda ,\nu ,\nu^\prime } a^{(\lambda)}_\nu
{a^{(\lambda )}_{\nu^\prime}}^*
\sum_{k ,k ^\prime} q_k q_{k ^\prime}
\left( M^{\{\lambda \} }_k M^{\{\lambda \} }_{k^\prime}
\right) _{\nu \nu^\prime}
\nonumber \\*
&\equiv & \frac{1}{2} \sum_{k , k^\prime} 
q_k q_{k^\prime} 
\mathcal{ J}_{k k^\prime} \ , \\
T_{\rm coup}&=& \frac{i}{2} \sum_{\lambda ,\nu ,\nu^\prime ,k} q_k 
\left[ \dot{a}^{(\lambda)}_\nu  {a^{(\lambda )}_{\nu^\prime}}^* -
a^{(\lambda)}_\nu  {{\dot{a}}^{(\lambda )*}_{\nu^\prime}} \right] 
\left( M^{\{ \lambda \} }_k \right) _{\nu ,\nu^\prime}\nonumber \\*
= &i&\!\sum_\lambda\!\sqrt{3 (2\lambda\!+\!1)} \left[ q^{(1)}\!\otimes\!\left[
a^{(\lambda )}\!\otimes \dot{a}^{(\lambda )} \right] ^{(1)} 
\right] ^{(0)}_0\!.
\label{E:a.6}
\end{eqnarray}
Here, $q_k$ ($k = 1,2,3$) are the Cartesian components of the angular
velocity $\vec q$ along the axes of the intrinsic frame, while 
the $M^{\{ \lambda \} }_k$ are $(2\lambda +1)$--dimensional matrices
giving the quantum--mechanical representation of the Cartesian 
components of  an angular momentum $M=\lambda$ in the intrinsic frame,
subject to commutation rules of the form
\begin{equation}
M^{\{ \lambda \} }_1M^{\{ \lambda \} }_2
- M^{\{ \lambda \} }_2M^{\{ \lambda \} }_1=-iM^{\{ \lambda \} }_3
\ . 
\label{E:a.7}
\end{equation}
We assume, as usually,
\begin{eqnarray*} 
&\left( \right. & M^{\{ \lambda \} }_3 \left. \right) _{\nu \nu^\prime}
= \nu \delta_{\nu ,\nu^\prime} \\*
&\left( \right.  & M^{\{ \lambda \} }_1 \left. \right) _{\nu \nu^\prime}
= \frac{1}{2}\ \left(  \sqrt{(\lambda - \nu)\ (\lambda + \nu + 1)}
\ \delta_{\nu^\prime,\nu +1}\right. \nonumber \\*
&&+\left. \sqrt{(\lambda + \nu)\ (\lambda - \nu + 1)}
\ \delta_{\nu^\prime,\nu -1}  \right) \\
&\left( \right. & M^{\{ \lambda \} }_2 \left. \right) _{\nu \nu^\prime}
= \frac{1}{2} \left( -i\ \sqrt{(\lambda - \nu)\ (\lambda + \nu + 1)}
\ \delta_{\nu^\prime,\nu +1} \right. \nonumber \\*
&&+\left. i\  \sqrt{(\lambda + \nu)\ (\lambda - \nu + 1)}
\ \delta_{\nu^\prime,\nu -1} \right) .
\label{E:a.7.1}
\end{eqnarray*}
Taking into account the properties of the $M^{\{ \lambda \} }_k$,
it is possible to obtain the explicit expression for the diagonal and
non diagonal elements of the tensor of inertia \cite{bohr52,wexler}
\begin{equation}
\mathcal{ J}_{k k^\prime} = \frac{1}{2}\sum_{\lambda, \nu, 
\nu^\prime}
a^{(\lambda)}_\nu {a^{(\lambda)}_{\nu^\prime}}^* \left ( M^{\{ \lambda \} }_k
 M^{\{ \lambda \} }_{k^\prime} 
+M^{\{ \lambda \} }_{k^\prime}  M^{\{ \lambda \} }_k
\right) _{\nu \nu^\prime}\ .
\label{E:a.7.2}
\end{equation}
The spin operators $M^{\{ \lambda \}}_k$ transform 
as the Cartesian components of a vector $\vec M^{\{ \lambda \}}$ under
rotation in the ordinary space.
 Taking into account
the commutation rule (\ref{E:a.7}), we can define the irreducible 
tensor components of  $\vec M^{\{ \lambda \}}$ as
\begin{eqnarray}
M^{(1)\{ \lambda \}}_0 &=& M^{\{ \lambda \}}_3 \nonumber \\
M^{(1)\{ \lambda \}}_{\pm 1} &=& \pm \frac{1}{\sqrt{2}}
\left( M^{\{ \lambda \}}_1
\mp i\ M^{\{ \lambda \}}_2 \right) 
\label{E:a.7b}
\end{eqnarray}
and express the products of two Cartesian components 
as the sum of products of two tensor components.
To this purpose, we
define, for each value of $\lambda$,
the irreducible tensor product
\begin{eqnarray}
T^{(J)}_m &=& \left[M^{(1)}\otimes M^{(1)}
\right] ^{(J)}_m \nonumber \\
&\equiv & \sum_{\nu ,\nu^\prime} (1 \nu 1 \nu^\prime |J m) 
M^{(1)}_\nu \ M^{(1)}_{\nu^\prime}
\end{eqnarray}
(where the common suffix $\{ \lambda\}$ has been dropped, 
for sake of simplicity). 
We now introduce the reduced matrix elements 
 of the tensor operator $T^{(J)}$, 
\begin{equation}
( \lambda ||T^{(J)} || \lambda )=(-1)^J\sqrt{2J+1}
\left\{\!\begin{array}{ccc}
1 & 1 & J \\
\lambda & \lambda & \lambda
\end{array}\!\right\} \lambda (\lambda +1)(2\lambda +1)
\end{equation}
to obtain
\begin{eqnarray}
\left( M^{(1)}_\mu \  M^{(1)}_{\mu^\prime}
\right) _{\nu ,\nu^\prime} &=& \sum_{Jm} (1 \mu 1 \mu^\prime | J m)
\ \left( T^{(J)}_m 
\right) _{\nu ,\nu^\prime} \\
&=&\sum_{Jm} (1 \mu 1 \mu^\prime | Jm) 
( \lambda ||T^{(J)} || \lambda )\nonumber \\
&&\times \frac{(-1)^{\lambda -\nu +m}}{\sqrt{2J+1}}
\ (\lambda \nu  \lambda -\nu^\prime | J m)\ . \nonumber
\end{eqnarray}
We now observe that 
\begin{equation}
{a^{(\lambda)}_{\nu^\prime}}^* = 
(-1)^{\nu^\prime} a^{(\lambda)}_{-\nu^\prime} 
\end{equation}
and that
\begin{equation}
\left[ a^{(\lambda)} \otimes a^{(\lambda)} \right]
^{(J)}_m
=\sum_{ \nu, \nu^\prime} a^{(\lambda)}_\nu 
a^{(\lambda)}_{-\nu^\prime}\  (\lambda \nu  \lambda -\nu^\prime | J m)
\end{equation}
to obtain
\begin{eqnarray}
\sum_{ \nu, \nu^\prime}&&
a^{(\lambda)}_\nu {a^{(\lambda)}_{\nu^\prime}}^* 
\left ( M^{(1)\{ \lambda \} }_\mu
 M^{(1)\{ \lambda \} }_{\mu^\prime} \right) _{\nu \nu^\prime}\nonumber
\\
&&= \sum_{J, m}
\frac{(-1)^{\lambda }}{\sqrt{2J+1}} (1 \mu 1 \mu^\prime | Jm)
\ ( \lambda ||
T^{(J)}
|| \lambda ) \nonumber \\
&&\ \times \ \left[ a^{(\lambda)} \otimes a^{(\lambda)} \right] ^{(J)}_m\ .
\end{eqnarray}
The rank of the tensor product $T^{(J)}$ of the two identical 
vectors $M^{(1)\{\lambda\}}$ must be even, and therefore the possible
values of $J$ are limited to 0 and 2.
Now we can substitute, in the eq.~(\ref{E:a.7.2}),
 the cartesian components of the
angular momentum with its tensor components
 defined in the eq.s (\ref{E:a.7b}).
The possible values of $m$ (and $J$) contributing to
the sum are limited to $m=0$ (and therefore $J=$0 or 2) in the case of
$(M_3 )^2$, to $m=0$ or 2 ($J=$0 or 2) for 
$(M_1 )^2$, $(M_2 )^2$ and
$(M_1 M_2 )$, to $m=1$ ($J=2$)
for $(M_3 M_1 )$ and $(M_3 M_2 )$.
One obtains
\begin{subequations}
\label{E:a.8}
\begin{eqnarray}
\mathcal{ J}_1 &=& \sum_\lambda  \left\{
 C_0(\lambda ) \left[  a^{(\lambda )} \otimes  
a^{(\lambda )} \right]^{(0)}_0\right. \nonumber \\*
&+& \frac{1}{\sqrt{6}} C_2(\lambda )
   \left[  a^{(\lambda )} \otimes  a^{(\lambda )} \right]^{(2)}_0
\nonumber \\*
&-& \left. C_2(\lambda ) {\rm Re}  
\left[  a^{(\lambda )} \otimes  a^{(\lambda )} 
\right]^{(2)}_2 \right\}\ ,\\
\mathcal{ J}_2 &=& \sum_\lambda \left\{
 C_0(\lambda ) \left[  a^{(\lambda )} \otimes  a^{(\lambda )}
\right]^{(0)}_0
\right. \nonumber \\*
&+& \frac{1}{\sqrt{6}} C_2(\lambda )
   \left[  a^{(\lambda )} \otimes  a^{(\lambda )} \right]^{(2)}_0
\nonumber \\*
&+&\left. C_2(\lambda ) 
{\rm Re}  \left[  a^{(\lambda )} \otimes  a^{(\lambda )} 
\right]^{(2)}_2 \right\} \ ,\\
\mathcal{ J}_3 &=& \sum_\lambda \left\{
 C_0(\lambda ) \left[  a^{(\lambda )} \otimes  a^{(\lambda )} 
\right]^{(0)}_0
\right. \nonumber \\*
&-& \sqrt{\frac{2}{3}}\left. C_2(\lambda )
   \left[  a^{(\lambda )} \otimes  a^{(\lambda )} \right]^{(2)}_0
 \right\} \ , \\
\mathcal{ J}_{12} &=& \sum_\lambda C_2(\lambda)
\ {\rm Im} \left[  a^{(\lambda )} \otimes  a^{(\lambda )}
\right]^{(2)}_2\ ,\\
\mathcal{ J}_{13} &=& \sum_\lambda  C_2(\lambda)
\ {\rm Re} \left[  a^{(\lambda )} \otimes  a^{(\lambda )}
\right]^{(2)}_1\ ,\\
\mathcal{ J}_{23} &=&  \sum_\lambda  C_2(\lambda)
\ {\rm Im} \left[  a^{(\lambda )} \otimes  a^{(\lambda )}
\right]^{(2)}_1 .
\end{eqnarray}
\end{subequations}
where
\begin{eqnarray*}
C_0(\lambda )&=& (\!-\!1)^\lambda \ \frac{ 
\lambda (\lambda+1)}{3} \sqrt{2 \lambda +1} \\
C_2(\lambda )&=&
(\!-\!1)^{\lambda +1}
\ \sqrt{\frac{\lambda  (\lambda \!+\!1)
(2\lambda \!+\!\!3)(4\lambda^2-1)}{30}} 
\end{eqnarray*}
and therefore
\begin{equation}
\begin{array}[c]{lp{8mm}l}
C_0(2)= 2\ \sqrt{5}\ , && C_0(3) = -4\ \sqrt{7}\ ,\\
C_2(2)=-\sqrt{21}\ , && C_2(3) = 3\ \sqrt{14}\ .
\end{array}
\label{E:a.9}
\end{equation}

If we chose as the intrinsic reference frame the principal axis of the
tensor of inertia, we must put $\mathcal{ J}_{12} =
\mathcal{ J}_{13} = \mathcal{
J}_{23} = 0$. Using Eq.s (\ref{E:1},\ref{E:2},\ref{E:3}) 
we obtain, up to the first order  
\begin{subequations}
\label{E:a.10}
\begin{eqnarray}
\mathcal{ J}_{12}&=& - 2\ \sqrt{21} \ {\rm Im}
\ \left[  \bar{a}^{(2)} \otimes  \tilde{a}^{(2)}
\right]^{(2)}_2
\nonumber \\[1pt]
&+& 6\ \sqrt{14}\ {\rm Im}
\ \left[  \bar{a}^{(3)} \otimes  \tilde{a}^{(3)} \right]^{(2)}_2 
= 0 ,\\
\mathcal{ J}_{13}&+& i \mathcal{ J}_{23} = - 2\ \sqrt{21}
\ \left[  \bar{a}^{(2)} \otimes  \tilde{a}^{(2)} \right]^{(2)}_1
\nonumber \\[1pt]
&+& 6\ \sqrt{14}
\ \left[  \bar{a}^{(3)} \otimes  \tilde{a}^{(3)} \right]^{(2)}_1 
 =0 \ .
\end{eqnarray}
\end{subequations}
Here we make use of the fact that the zero--order terms are
automatically set to zero if the $\bar{a}^{(\lambda )}_\mu$ are defined
according to eq.s (\ref{E:1},\ref{E:2}). By inserting these definitions
in the above equations, and retaining only the first--order  terms, 
one obtains
\begin{eqnarray*}
\left[  \bar{a}^{(2)} \otimes  \tilde{a}^{(2)}  \right]^{(2)}_1
&\approx &   (2,0,2,1|2,1)\  \beta_2\ \tilde{a}^{(2)}_1 \\
&=&  -\frac{1}{\sqrt{14}}\ \beta_2\ \tilde{a}^{(2)}_1 \ ,
\\
{\rm Im} \left[  \bar{a}^{(2)} \otimes  \tilde{a}^{(2)}  \right]^{(2)}_2
&\approx  &  (2,0,2,2|2,2)\   \beta_2\ {\rm Im}\ \tilde{a}^{(2)}_2\cr
&=&\sqrt{\frac{2}{7}}\  \beta_2\ {\rm Im}\ \tilde{a}^{(2)}_2 \ ,
\\
\left[  \bar{a}^{(3)} \otimes  \tilde{a}^{(3)}  \right]^{(2)}_1
&\approx &   (3,0,3,1|2,1)\ \beta_3\ \tilde{a}^{(3)}_1\cr
&=&\sqrt{\frac{1}{42}}\ \beta_3\ \tilde{a}^{(3)}_1 
,\\
{\rm Im} \left[  \bar{a}^{(3)} \otimes  \tilde{a}^{(3)}  \right]^{(2)}_2
&\approx & (3,0,3,2|2,2)\   \beta_3\ {\rm Im}\ \tilde{a}^{(3)}_2 \cr
&=& -\sqrt{\frac{5}{21}}\  \beta_3\ {\rm Im}\ \tilde{a}^{(3)}_2 .
\end{eqnarray*}
We obtain, therefore, from eq.s~(\ref{E:a.10})
\label{E:a.12}
\begin{eqnarray}
\beta_2 \tilde{a}^{(2)}_1 &=& - \sqrt{2}\ \beta_3 \tilde{a}^{(3)}_1 ,
\nonumber \\*
\beta_2 {\rm Im}\ \tilde{a}^{(2)}_2 &=&  
- \sqrt{5}\ \beta_3  {\rm Im}\ \tilde{a}^{(3)}_2 .
\end{eqnarray}

\section{\label{S:a2} Effect of non diagonal (first-order) terms}

The matrix $\mathcal{ G}$, as it results from the Table~\ref{T:1},
contains zero-order terms only in its principal diagonal (whose
last element, however, is small of the second order). Non diagonal
terms
have been expanded in series up to the first order in the ``small''
dynamical variables (all of them, apart from $\beta_2$ and $\beta_3$).
We shall call $\xi_k$ a generic ``small'' variable, different from
$\beta_\lambda$\ $(\lambda=2,3)$.

Terms of the last line and column (those related to the third intrinsic
component of the angular velocity) are discussed in the 
Section~\ref{S:2.4}. 
Here, we consider a simpler problem: the inversion of a
matrix $\mathcal{ G}$ which has finite values for all terms 
in the principal
diagonal, and only ``small'' values for all others. 
In the zero-order approximation, the inverse matrix 
$\mathcal{ A = G}^{-1}$
is diagonal, with diagonal elements $A_{\mu \mu} = 1/ G_{\mu \mu}$.

The first--order approximation gives the non--diagonal elements
\begin{equation}
A_{\mu \nu} = A_{\nu \mu} = -\frac{G_{\mu \nu}}{G_{\mu \mu} 
G_{\nu \nu}} 
\label{E:nd1}
.\end{equation}
We are interested in particular on the effect of non diagonal terms on
the coefficients of the derivatives with respect to $\beta_2$ or
$\beta_3$. 

Terms involving the derivatives with respect to
$\beta_\lambda$ and $\xi_k$ have the form

\begin{equation}
G^{-\frac{1}{2}} \frac{\partial }{\partial \xi_k}
\left[ G^\frac{1}{2} A_{k \lambda} \frac{\partial }{\partial 
\beta_\lambda} \right]
+G^{-\frac{1}{2}} \frac{\partial }{\partial \beta_\lambda}
\left[ G^\frac{1}{2} A_{\lambda k} \frac{\partial }{\partial \xi_k} 
\right] .
\label{E:nd2}
\end{equation}
The non diagonal matrix elements of the matrix $\mathcal{ A}$ are of the
first order in the ``small'' variables. They have therefore  the form 
$f^{(i)}_{\lambda k}(\beta_2,\beta_3) \xi_i$, where 
$f^{(i)}_{\lambda k}(\beta_2,\beta_3)=
\partial \mathcal{ A}_{\lambda ,k} /\ \partial \xi_i$, 
or -- possibly -- are the sum of
several terms like that. 

Substituting this expression in the
Eq.~(\ref{E:nd2}) one obtains
\begin{eqnarray}
&G&^{-\frac{1}{2}} \frac{\partial }{\partial \xi_k} 
\left[ G^\frac{1}{2} A_{k \lambda} 
\frac{\partial }{\partial \beta_\lambda} \right]
+ G^{-\frac{1}{2}} 
\frac{\partial }{\partial \beta_\lambda}
\left[ G^\frac{1}{2} A_{\lambda k} \frac{\partial }{\partial \xi_k} 
 \right] \nonumber \\[1pt]
&=& f^{(i)}_{\lambda k}(\beta_2,\beta_3)\left[ 2 \xi_i 
\frac{\partial }{ \partial \xi_k}
+ \frac{\xi_i}{2 G}\ \left( \frac{\partial G}{\partial \xi_k} \right)
+ \delta_{ik}  \right] \frac{\partial }{\partial \beta_\lambda} 
\nonumber \\*
&+& \xi_i \left[ f^{(i)}_{\lambda k}(\beta_2,\beta_3)
\frac{1}{2 G}\ \left( \frac{\partial G}
{\partial \beta_\lambda} \right)  
+\left( \frac{\partial f^{(i)}_{\lambda k} 
}{\partial \beta_\lambda} \right) \right] \frac{\partial }{\partial
\xi_k}\ .
\label{E:nd3}
\end{eqnarray}
The last line of Eq.~(\ref{E:nd3}) only contains the partial-derivative
operator with respect to the small variable $\xi_k$. 
Compared with the diagonal term involving the corresponding 
second-derivative operator, it contains a
small factor $\xi_i$ more, and can be neglected. 

In the first line
of the expression, all terms are potentially of the same order of 
magnitude of
the leading ones, and in principle \textsl{ could not} be neglected.
The first of the terms in the square brackets 
contains  the partial derivative with respect 
to $\xi_k$. In the spirit of the adiabatic approach, we try to
estimate the expectation value of this term for the ground--state 
wavefunction in all the variables, with the exception of $\beta_2$ and
$\beta_3$. 
 If we assume that the potential is harmonic for the ensemble of these
variables, the ground--state wavefunction can be expected to be close
to a multivariate Gaussian function. If, moreover, $\xi_k$ 
and $\xi_i$ are
uncorrelated, the expectation value of $\xi_i \partial / \partial \xi_k$
is zero, as $\left< \xi_i \right> =0 $.

For the case $i=k$, instead, the expectation value is
\begin{eqnarray}
\left< 2\xi \frac{\partial}{\partial \xi} \right> &\approx& 
2\ \frac{ \int  \exp (-\alpha \xi^2/2) \xi 
\frac{\partial  }{\partial  \xi} \exp (-\alpha \xi^2/2) {\rm d}\xi}
{\int   \exp (-\alpha \xi^2) {\rm d}\xi }\nonumber \\[1pt]
&=&\frac{\int \exp (-\alpha \xi^2/2) \left( -2\alpha \xi^2 \right) 
 \exp (-\alpha \xi^2/2) {\rm d}\xi}
{\int  \exp (-\alpha \xi^2) {\rm d}\xi }\cr
&=&-1
\label{E:nd4}
\end{eqnarray}
and (in this approximation) cancels the third term, $\delta_{ik}$.
In conclusion, the coefficient of $\partial  / \partial \beta_\lambda$
coming from the non--diagonal terms of the matrix ${\mathcal G}$
 can be approximated with a sum of expressions like
\begin{equation}
\ f^{(i)}_{\lambda k}(\beta_2,\beta_3)\ \left< \frac{\xi_i}{G} 
\ \frac{\partial G}{\partial \xi_k} \right>
\label{E:nd5}
\end{equation}
which approximately vanishes for $i\neq k$, and also
vanishes for $i=k$ unless both
$\mathcal{ A}_{\lambda k}$ and $G$ depend explicitly on $\xi_k$.
Moreover, also in this latter case,
it is possible to eliminate non-diagonal elements of the matrix
 $\mathcal{ G}$ of the form  $\mathcal{ G}_{\lambda k}=
\ g_{\lambda k}(\beta_2,\beta_3)\ \xi_k$, with a slight
change in the definition of $\beta_\lambda $, without any other
effect at the present order of approximation.
Namely, it is sufficient to substitute the dynamical variable
$\beta_\lambda$ with the new variable
\begin{equation}
\beta^{\rm o}_\lambda=\beta_\lambda+\frac{1}{2}g_{\lambda k}
(\beta_2,\beta_3)  \xi_k^2
\label{E:nd6}
\end{equation}
to obtain
\begin{equation}
\dot{\beta}_\lambda=\dot{\beta}^{\rm o}_\lambda
-g_{\lambda k}(\beta_2,\beta_3) \xi_k \dot{\xi}_k
+ 0(\xi_k^2)
\end{equation}
and therefore, up to the second order in $\xi_k$,
\begin{eqnarray}
&\dot{\beta}_\lambda^2& + 2 g_{\lambda k}(\beta_2,\beta_3)\xi_k
\dot{\beta}_\lambda \dot{\xi}_k\nonumber \\*
&=&
\left( \dot{\beta}^{\rm o}_\lambda
-g_{\lambda k}(\beta_2,\beta_3) \xi_k \dot{\xi}_k \right) ^2
+ 2 g_{\lambda k}(\beta_2,\beta_3)\xi_k
\dot{\beta}_\lambda \dot{\xi}\nonumber \\*
&\approx& \left( \dot{\beta}^{\rm o}_\lambda
\right) ^2 .
\end{eqnarray}
This argument can be easily extended to the case in which the
dependence of $ \mathcal{ A}_{\lambda k}$ on $\xi_k$ comes from the
dependence on  $\xi_k$ of one of the terms at the denominator in the 
eq.~(\ref{E:nd1}).

\section{\label{S:a4} Quantization according to the Pauli rule}
\begingroup
\squeezetable
{\renewcommand{\arraystretch}{1.30}
\begin{table*}[t]
\caption{\label{AT:1}  The relevant part of the  matrix of inertia,
 $\mathcal{ G}_1^\prime$, 
after the substitution of the angular-velocity components $q_k$
($k=1,2,3$) with the time derivative of the Euler angles $\theta_k$.
Here, $\mathcal{ J}_1 = \mathcal{ J}_2 = 3(\beta_2^2 + 2\beta_3^2)$, and
$\mathcal{ J}_3 = 4 \ u_0^2 + 
2v^2 + 8u^2 + 18w^2$.
The determinant of this $6\times 6$  matrix is now $G_1^\prime
 = {\rm Det}\ \mathcal{ G}_1^\prime =
288\ u_0^2\ v^2 u^2 w^2 
\big( \beta_2^2 + 2 \beta_3^2 \big) ^2\ \sin^2 \theta_2$ .
}
\begin{ruledtabular}
\begin{tabular}{c|ccc|ccc}
& & & & & &\\[-9pt]
& $\dot{\varphi} $ & $\dot{\chi}$ &  $\dot{\vartheta}$ &
$\dot{\theta}_1$ & $\dot{\theta}_2$ & $\dot{\theta}_3$ \\[2pt]
\hline 
$\dot{\varphi}$ & $2v ^2$   & 0 & 0 & $2v ^2\cos \theta_2$ & 0 &
                                           $2v ^2$ \\
$\dot{\chi}$    & 0 & $2u^2$  & 0 &  $4u^2\cos \theta_2$
   & 0  &      $4u^2$ \\
$\dot{\vartheta}$ & 0 & 0& $2 w^2$ & $6 w^2\cos \theta_2$  & 0  
  & $6 w^2$  \\
\hline 
& & & & & &\\[-9pt]
$\dot{\theta}_1$ & 
 $2v ^2\cos \theta_2$ &
$4u^2\cos \theta_2$ &
$6 w^2\cos \theta_2$ &
{\renewcommand{\arraystretch}{1.0}
 $\begin{array}{c}(\mathcal{  J}_1\cos^2\theta_3 +\mathcal{
J}_2\sin^2\theta_3)\\ 
*\sin^2\theta_3 
+\mathcal{ J}_3 \cos^2\theta_2
\end{array}$} &
{\renewcommand{\arraystretch}{1.0}
$\begin{array}{c}(\mathcal{  J}_2 -\mathcal{ J}_1)\sin \theta_2\\ 
*\sin \theta_3  \cos\theta_3
\end{array}$} 
 &  $\mathcal{ J}_3  \cos^2 \theta_2
 $ \\[8pt]
$\dot{\theta}_2$ & 0 & 0 & 0 & 
{\renewcommand{\arraystretch}{1.}
$\begin{array}{c}(\mathcal{  J}_2 -\mathcal{ J}_1)\sin \theta_2\\ 
*\sin \theta_3  \cos\theta_3
\end{array}$} &
$\mathcal{ J}_1 \sin^2\theta_3+\mathcal{ J}_2 \cos^2\theta_3$ & 0 
\\[8pt]
$\dot{\theta}_3$ & $2v^2$ & $4u^2$ & $6w^2$ & 
$\mathcal{ J}_3  \cos^2 \theta_2$ & 0 
 & $\mathcal{ J}_3$ \\[2pt]
\end{tabular}
\end{ruledtabular}
\end{table*}
}
\endgroup
In this Appendix, we discuss some aspects of the quantization of the
kinetic energy expression given, \textit{ e.g.}, in the 
eq.~(\ref{E:6.51c}),
by means of the Pauli procedure: namely, if 
the classical expression of the kinetic energy in terms of the time
derivatives of the dynamical variables $\xi_\mu$ is
\begin{equation}
T=\frac{1}{2}\sum_{\mu ,\nu}\mathcal{ G}_{\mu ,\nu} 
\dot{\xi}_\mu \dot{\xi}_\nu
\label{EA:1}
\end{equation}
the corresponding quantum operator has the form
\begin{equation}
\hat T=-\frac{\hbar^2}{2} G^{-1/2}\sum_{\mu ,\nu}
\frac{\partial}{\partial \xi_\mu}\ G^{1/2} \mathcal{A}_{\mu ,\nu} 
\frac{\partial}{\partial \xi_\nu}
\label{EA:2}
\end{equation}
where $G={\rm Det}\ \mathcal{ G}$ and $\mathcal{ A}=\mathcal{ G}^{-1}$.
 The choice of the
``best set'' of
 dynamical variables is, in part, related to the expression of the
potential-energy term, and it is not obvious that it will eventually
coincide with the one discussed in the section \ref{S:2.5}. However, it
can be useful to explore the properties of the kinetic-energy
operator in the particular model in which the matrix of coefficients
${\mathcal G}$ is exactly that of the Table~\ref{T:2}, with the
non-diagonal terms confined in one single line (and column) of the
lower $6\times 6$ sub-matrix. 
{\renewcommand{\arraystretch}{1.50}
\renewcommand{\tabcolsep}{9pt}
\begin{table*}[ht]
\caption{\label{AT:3} The matrix 
$i (\mathcal{ G}_1^\prime )^{-1} \widetilde{W} \sin \theta_2$
(see  text). }
\begin{ruledtabular}
\begin{tabular}{l|ccc|ccc}
& $\hat{p}_{\varphi} $ & $\hat{p}_{\chi}$ &  $\hat{p}_{\vartheta}$ &
$\hat{L}_1$ & $\hat{L}_2$ & $\hat{L}_3$ \\
\hline
$\frac{\partial}{\partial \varphi}$ & 
$\frac{\sin \theta_2}{2v ^2} + \frac{\sin \theta_2}{4u_0^2}$
& $\frac{2\sin \theta_2}{4u_0^2}$ & $\frac{3\sin 
\theta_2}{4u_0^2}$ & 0 & 0
& $-\frac{\sin \theta_2}{4u_0^2}$ \\
$\frac{\partial}{\partial \chi}$ & $\frac{2\sin 
\theta_2}{4u_0^2}$ & 
$\frac{\sin \theta_2}{2u^2}+\frac{4\sin \theta_2}{4u_0^2}$ & 
$\frac{6\sin \theta_2}{4u_0^2}$
& 0 & 0 & $-\frac{2\sin \theta_2}{4u_0^2}$\\
$\frac{\partial}{\partial \vartheta}$ 
&  $\frac{3\sin \theta_2}{4u_0^2}$ &  
$\frac{6\sin \theta_2}{4u_0^2}$
  & $\frac{\sin \theta_2}{2 w^2} + \frac{9\sin 
\theta_2}{4u_0^2}$ & 0 & 0 & $-\frac{3\sin 
\theta_2}{4 u_0^2}$ \\[3pt]
\hline
$\frac{\partial}{\partial \theta_1}$ & 0 & 0 & 0 & 
$-\frac{\cos \theta_3}{\mathcal{ J}_1}$ &  $\frac{\sin 
\theta_3}{\mathcal{ J}_2}$  & 0 \\
$\frac{\partial}{\partial \theta_2}$ & 0 & 0 & 0 & 
 $\frac{\sin \theta_2 \sin \theta_3}{\mathcal{ J}_1}$  
& $\frac{\sin \theta_2 \cos \theta_3}{\mathcal{ J}_2}$
 & 0 \\
$\frac{\partial}{\partial \theta_3}$ & 
$-\frac{\sin \theta_2}{4u_0^2}$ & $ -\frac{2\sin 
\theta_2}{4u_0^2} $ 
& $-\frac{3\sin \theta_2}{4u_0^2}$ &
$\frac{\cos \theta_2 \cos \theta_3}{\mathcal{ J}_1}$
& $-\frac{\cos \theta_2 \sin \theta_3}{\mathcal{ J}_2}$ &
 $\frac{\sin \theta_2}{4u_0^2}$  \\[3pt]
\end{tabular}
\end{ruledtabular}
\end{table*}
}

From now on, we  limit our
discussion to the lowest $6\times 6$ sub-matrix $\mathcal{ G}_1$ of the
matrix $\mathcal{ G}$ (and of the matrices derived from $\mathcal{G}$,
Tables \ref{AT:3} ,\ref{AT:2}) as
the six corresponding  variables -- $\varphi ,\ \chi ,\ \vartheta
,\ \theta_1,\ \theta_2,\ \theta_3$ -- are effectively decoupled from
the others.

The formal procedure is discussed, \textit{ e.g.}, in the Ch.s 5 and 
6 of the ref.~\cite{eisen1}.
The first pass is the substitution of the intrinsic components 
$q_1,\ q_2,\ q_3$ of the angular velocity with the time derivatives of
the three Euler angles $\theta_1,\ \theta_2,\ \theta_3$: 
\begin{equation}
q_k=\sum_i V_{ki} \dot{\theta}_i
\label{EA:3}
\end{equation}
with
\begin{equation}
V=\left| \begin{array}{ccc}
-\cos \theta_3 \sin \theta_2 & \sin \theta_3 & 0  \\
\phantom{-}\sin \theta_3 \sin \theta_2 &  \cos \theta_3 & 0 \\
\cos \theta_2  &  0 & 1\\
\end{array}
\right| \ .
\label{EA:4}
\end{equation}
As a consequence of this substitution, the $6\times 6$ matrix 
$\mathcal{ G}_1$ transforms according to the relation
\begin{equation}
\mathcal{ G}_1^\prime = \widetilde{W}\ 
\ \mathcal{ G}_1\  W,\hspace{20mm} W= \left| \begin{array}{cc}
\{1\} & \{0\} \\
\{0\} & V \end{array}
\right|
\label{EA:5}
\end{equation}
(where $\widetilde{W}$ is the transpose of the matrix $W$ and $\{1\}$,
$\{0\} $ are the $3\times 3$ unit matrix and null matrix) and 
 takes the form shown in the  Table \ref{AT:1}.
{\renewcommand{\arraystretch}{1.30}
\renewcommand{\tabcolsep}{4.5pt}
\begin{table}[ht]
\caption{\label{AT:2} The lowest $6\times 6$ sub-matrix of the 
inverse of the matrix $\mathcal{ G}$ given in the Table~\ref{T:2}. 
The upper part is diagonal in the present approximation.}
\begin{ruledtabular}
\begin{tabular}{c|ccc|ccc}
& $\hat{p}_{\varphi} $ & $\hat{p}_{\chi}$ &  $\hat{p}_{\vartheta}$ &
$\hat{L}_1$ & $\hat{L}_2$ & $\hat{L}_3$ \\[3pt]
\hline
$\hat{p}_{\varphi}$ & $\frac{1}{2v ^2} + \frac{1}{4u_0^2}$
& $\frac{2}{4u_0^2}$ & $\frac{3}{4u_0^2}$ & 0 & 0
& $-\frac{1}{4u_0^2}$ \\
$\hat{p}_{\chi}$ & $\frac{2}{4u_0^2}$ & 
$\frac{1}{2u^2}+\frac{4}{4u_0^2}$ & $\frac{6}{4u_0^2}$
& 0 & 0 & $-\frac{2}{4u_0^2}$\\
$\hat{p}_{\vartheta}$ &  $\frac{3}{4u_0^2}$ &  
$\frac{6}{4u_0^2}$
  & $\frac{1}{2 w^2} + \frac{9}{4u_0^2}$ & 0 & 0 & 
$-\frac{3}{4u_0^2}$ \\[3pt]
\hline
$\hat{L}_1$ & 0 & 0 & 0 & $\frac{1}{\mathcal{ J}_1}$ & 0 & 0 \\
$\hat{L}_2$ & 0 & 0 & 0 & 0 & $\frac{1}{\mathcal{ J}_2}$ & 0 \\
$\hat{L}_3$ &$-\frac{1}{4u_0^2}$ & $ -\frac{2}{4u_0^2} $ 
& $-\frac{3}{4u_0^2}$ & 0 & 0 & $\frac{1}{4u_0^2}$  \\[5pt]
\end{tabular}
\end{ruledtabular}
\end{table}
}
The next step is the inversion of the matrix $\mathcal{G}_1^\prime$. 
In doing this, the explicit form (\ref{E:6.51b}) of $\mathcal{ J}_3$
 has  been introduced
and second-order terms in the small amplitudes have been
neglected.
At this point, it is necessary to introduce the intrinsic components
of the angular momentum operator ($L_1,\ L_2,\ L_3$) in the place of
the derivatives with respect to the Euler angles. The expression of
$L_k$ in terms of $\partial /\partial \theta_k$ and vice-versa are 
given, \textit{ e.g.} in the Ch. 5 of ref~\cite{eisen1}. One gets
\begin{equation}
-i \frac{\partial}{\partial \theta_k} = \sum V_{ik} \hat{L}_i
\end{equation}
with the matrix $V$ given by the eq.~(\ref{EA:4}).
It is also convenient to  define the quantum operators 
\begin{eqnarray}
\hat{p}_\varphi &=& -i\partial \ / \ \partial \varphi\nonumber \\*
\hat{p}_\chi &=& -i \partial \ /\ \partial \chi\nonumber \\*
\hat{p}_\vartheta &=& -i\partial\ /\ \partial \vartheta
\label{EA:4a}
\end{eqnarray}
to obtain
\begin{eqnarray}
&-i& \left\{ \frac{\partial}{\partial \varphi},
\ \frac{\partial}{\partial \chi},
\ \frac{\partial}{\partial \vartheta},
\ \frac{\partial}{\partial \theta_1},
\ \frac{\partial}{\partial \theta_2},
\ \frac{\partial}{\partial \theta_3} \right\} \nonumber \\[1pt]
&&=\ \widetilde{W}\ \left\{ \hat{p}_\varphi ,\ \hat{p}_\chi ,
\ \hat{p}_\vartheta ,
\ \hat{L}_1,\ \hat{L}_2,\ \hat{L}_3 \right\} .
\label{EA:4b}
\end{eqnarray}

The introduction of a set of conjugate momenta $\hat{p}_k$ such that
\begin{equation}
 -i \frac{\partial}{\partial \xi_k} = \sum_i W_{ik} \hat{p}_i  
\label{EA:5a}
\end{equation}
in the eq.~(\ref{EA:2}) gives
\begin{eqnarray}
\hat T&=&\frac{\hbar^2}{2}\left[ \sum_{\mu ,\nu} 
(W \mathcal{ A} \widetilde{W})_{\mu ,\nu}\ \hat{p}_\mu \hat{p}_\nu \right. 
\nonumber \\*
&&\left.  +i\ G^{-1/2}\sum_{\mu ,\nu} 
\frac{\partial\ G^{1/2} (\mathcal{ A} \widetilde{W})_{\mu ,\nu}}{\partial 
\xi_\mu}\ \hat{p}_\nu
\right] .
\label{EA:6}
\end{eqnarray}
In our case, $\mathcal{ A}= (\mathcal{ G}_1^\prime )^{-1}$, 
$\mathcal{ G}_1^\prime = \widetilde{W} \mathcal{ G}_1 W$, so that
\begin{equation}
W \mathcal{ A}\widetilde{W} = W \left( W^{-1} 
\mathcal{ G}_1^{-1} \widetilde{W}^{-1} \right)
\widetilde{W} = \mathcal{ G}_1^{-1}
\label{EA:7}
\end{equation}
and the first term of the eq. (\ref{EA:6}) takes the form given in the
Table~\ref{AT:2}. 
The second term of the eq. (\ref{EA:6}) vanishes. 
In fact, only the derivatives with respect to $\theta_2$ or
$\theta_3$ could give a contribution to the sum, as the other
variables do not appear in the elements of the sub-matrix. Moreover,
the determinant $G_1$ is simply proportional to $\sin^2 \theta_2$
(with the proportionality factor depending on dynamical variables
which are outside the present sub-space), and
the term $G^{1/2}$ in the equation (\ref{EA:6}) can be replaced by 
$\sin \theta_2$.
The matrix $ i (\mathcal{ G}_1^\prime )^{-1} \widetilde{W} \sin \theta_2$ 
is given in the
Table~\ref{AT:3}. The sum of the derivatives of each element of 
the 5th row, with respect to $\theta_2$, plus the corresponding one of
the 6th row, with respect to $\theta_3$, would give the coefficient of
the corresponding momentum operator, but
it is easy to verify they cancel each other.
\begingroup
\squeezetable
\ifthenelse{\linewidth = \textwidth}{\begin{turnpage} }{}
{\renewcommand{\arraystretch}{1.80}
\begin{table*}[ht]
\caption{\label{T:3} Invariants up to the fourth order built with
the tensors $a^{(2)}$ and $a^{(3)}$, in the limit close to axial
symmetry. 
Here $T^{(L)}_{KK} \equiv \big[ a^{K} \otimes  a^{K} \big] ^{(L)}$,
and we use the definitions $v^2 = \eta^2 + \zeta^2$ ,
$w^2=X^2+Y^2$. 
``Small'' (non-axial)  terms are approximated up to the
second order.}
\begin{ruledtabular}
\begin{tabular}{lcccccc}
$A_1\equiv $ &
$\big( a^{(2)} \cdot a^{(2)}\big) =$ & 
$\beta_2^2 $ & & &
$ +\frac{10 \beta_3^2 }{\beta_2^2+5\beta_3^2}\xi^2$ & 
$+\frac{4\beta_3^2 }{\beta_2^2+2\beta_3^2} v^2$
\\
$A_2\equiv $ &
$ -\big( a^{(3)} \cdot a^{(3)}\big)  =$ &
$ \beta_3^2 $ & & $+10 w^2 $ &
$ +\frac{2 \beta_2^2 }{\beta_2^2+5\beta_3^2}\xi^2$ & 
$+\frac{2 \beta_2^2  }{\beta_2^2+2\beta_3^2}v^2
$
\\
$B_1\equiv $ &
$ - \frac{\sqrt{14}}{4}\ \big( a^{(2)} \cdot T^{(2)}_{22}\big)  =$ &
$ \beta_2^3$ & 
$-\frac{9}{2} \beta_2^3 \gamma_2^2 $& &
$-\frac{30 \beta_2 \beta_3^2 }{\beta_2^2+5\beta_3^2}\xi^2 $ &
$+\frac{6  \beta_2 \beta_3^2  }{\beta_2^2+2\beta_3^2}v^2$ 
\\  
$B_2\equiv $ &
$\frac{\sqrt{21}}{2}\ \big( a^{(2)} \cdot T^{(2)}_{33} \big)  =$ &
$ \beta_2 \beta_3^2$ &
$ -\beta_2 \beta_3^2\ \big( \frac{1}{2} \gamma_2^2
+\sqrt{5} \gamma_2 \gamma_3 + \gamma_3^2 \big) $ &
$-\frac{25}{2} \beta_2 w^2   $ &
$+\frac{10\beta_2 \beta_3^2 }{\beta_2^2+5\beta_3^2}\xi^2$  &
$\frac{3\beta_2^3-4\beta_2 \beta_3^2 }{2(\beta_2^2+2\beta_3^2)}v^2 $
\\
$C_1\equiv $ &
$ \frac{7}{2}\ \big(T^{(2)}_{22} \cdot T^{(2)}_{22} \big)  =$ &
$\beta_2^4$ & & & 
$+\frac{20 \beta_2^2 \beta_3^2 }{\beta_2^2+5\beta_3^2}\xi^2 $ & 
$+\frac{8  \beta_2^2 \beta_3^2  }{\beta_2^2+2\beta_3^2}v^2 $ 
\\
$C_2\equiv $ &
$\frac{35}{18}\ \big( T^{(4)}_{22} \cdot T^{(4)}_{22} \big)  =$ &
$\beta_2^4$ & & & 
$+\frac{20 \beta_2^2 \beta_3^2 }{\beta_2^2+5\beta_3^2}\xi^2 $ &
$+\frac{8  \beta_2^2 \beta_3^2  }{\beta_2^2+2\beta_3^2}v^2 $ 
\\
$C_3\equiv $ &
$\frac{21}{4}\ \big( T^{(2)}_{33} \cdot T^{(2)}_{33} \big)  =$ &
$\beta_3^4 $ &
$+3\beta_3^4 \gamma_3^2 $ &
$-25\beta_3^2  w^2  $ &
$+\frac{10 \beta_2^2 \beta_3^2 }{\beta_2^2+5\beta_3^2}\xi^2 $ &
$+ \frac{4\beta_2^2 \beta_3^2  }{\beta_2^2+2\beta_3^2}v^2 $
\\
$C_4\equiv $ &
$\frac{77}{18}\ \big( T^{(4)}_{33} \cdot T^{(4)}_{33} \big)  =$ &
$\beta_3^4 $ &
$ -4 \beta_3^4 \gamma_3^2 $ &
$ +80 \beta_3^2  w^2  $ &
$ -\frac{4 \beta_2^2 \beta_3^2 }{\beta_2^2+5\beta_3^2}\xi^2 $ &
$ + \frac{4 \beta_2^2 \beta_3^2   }{\beta_2^2+2\beta_3^2}v^2 $
\\
$C_5\equiv $ &
$ \frac{231}{100}\ \big( T_{33}^{(6)} \cdot T^{(6)}_{33} \big)  =$ &
$\beta_3^4 $ &
$+\frac{21}{25} \beta_3^4 \gamma_3^2 $ &
$+\frac{37}{5}\beta_3^2  w^2  $ &
$+\frac{142 \beta_2^2 \beta_3^2 }{25(\beta_2^2+5\beta_3^2)}\xi^2 $ &
$+\frac{4\beta_2^2 \beta_3^2  }{\beta_2^2+2\beta_3^2}v^2 $
\\
$C_6\equiv $ &
$-\frac{7\sqrt{6}}{4} \big( T^{(2)}_{22} \cdot T^{(2)}_{33} \big)  =$ &
$\beta_2^2 \beta_3^2 $ &
$- \beta_2^2 \beta_3^2 \big(2 \gamma_2^2 -2\sqrt{5} \gamma_2 \gamma_3
+ \gamma_3^2\big)$ &
$-\frac{25}{2} \beta_2^2 w^2  $ &
$-\frac{20 \beta_2^2 \beta_3^2 
+10 \beta_3^4}{\beta_2^2+5\beta_3^2}\xi^2 $ &
$+\frac{3\beta_2^4 -4 \beta_2^2 \beta_3^2 +4 \beta_3^4 
 }{2(\beta_2^2+2\beta_3^2)}v^2 $
\\
$C_7\equiv $ &
$-\frac{7\sqrt{55}}{18}\big( T^{(4)}_{22} \cdot T^{(4)}_{33} \big)  =$ &
$\beta_2^2 \beta_3^2 $ &
$-\frac{1}{6}\beta_2^2 \beta_3^2 \big(5 \gamma_2^2 +2\sqrt{5} 
\gamma_2 \gamma_3
+13 \gamma_3^2\big)$ &
$+5 \beta_2^2 w^2  $ &
$+\frac{-7\beta_2^4+ 10 \beta_2^2 \beta_3^2 
+5 \beta_3^4  }{3( \beta_2^2+5\beta_3^2)}\xi^2 $ &
$+\frac{\beta_2^4 -20 \beta_2^2 \beta_3^2 -8 \beta_3^4 
 }{3(\beta_2^2+2\beta_3^2)}v^2$
\\[3pt]
\end{tabular}
\end{ruledtabular}
\end{table*}
}
\ifthenelse{\linewidth =\textwidth}{\end{turnpage} }{}
\endgroup
\section{\label{S:a3} The possible forms of the potential term}

The form of the approximate potential--energy expression depends on 
the details of the underlying microscopic structure that the model 
should
try to simulate. There are, however, some general rules to which the
expression of the potential energy must conform: it must be invariant
under space rotation, time reversal, and parity.

Irreducible tensors of different rank have been constructed with each of
the basic tensors $a^{(2)}$ and $a^{(3)}$, and scalar products of
tensors of equal rank have been considered. 
In order to be invariant under time reversal, each term must
contain an even number of octupole amplitudes $a^{(3)}_\mu$.
Moreover, due to symmetry, only tensors 
$T^{(K)}_{ \lambda \lambda} = 
[a^{( \lambda )}\otimes a^{( \lambda )}]^{(K)}$
of even rank $K$ can be obtained
with the coupling of two identical tensors $a^{( \lambda )}$. 
There are therefore two independent invariants of order 2, 
$\big( a^{(2)} \cdot a^{(2)}\big)$ and $\big( a^{(3)} 
\cdot a^{(3)}\big)$,
two of order 3, $\big( a^{(2)} \cdot T^{(2)}_{22}\big)$
and $\big( a^{(2)} \cdot T^{(2)}_{33}\big)$,
 and seven of order 4, namely 
 $\big( T^{(K)}_{22} \cdot T^{(K)}_{22}\big)$ ($K=2,4$), 
 $\big( T^{(K)}_{33} \cdot T^{(K)}_{33}\big)$ ($K=2,4,6$), 
 and  $\big( T^{(K)}_{22} \cdot T^{(K)}_{33}\big)$ ($K=2,4$).
 Additional fourth-order invariants of the form 
 $\big( [a^{(2)}\otimes a^{(3)}]^{(K)} \cdot [a^{(2)} \otimes
 a^{(3)}]^{(K)}\big)$ ($K=1-5$), are not independent from the above
ones, and can be expressed as linear combination of them with the
standard rules of angular momentum recoupling. 

Invariant expressions up to the fourth order in the amplitudes 
$a^{(\lambda )}_\mu$ ($\lambda =2,3$) are shown in the
Table~\ref{T:3}, in terms of the dynamical variables 
$\beta_2$, $\gamma_2$, $\beta_3$, $\gamma_3$, $w^2=(X^2+Y^2)$,
$\xi$, and $v^2=(\eta^2 +\zeta^2)$. Expressions corresponding to
different choices of the dynamical variables can be easily obtained.

Here, only the terms up to the second order in the series expansion of
the ``small'' (non-axial) amplitudes are given. In this approximation,
the fourth-order invariants built with the quadrupole amplitudes 
($C_1,\ C_2$)
result to be proportional to each other and to the square of the 
corresponding second-order invariant ($A_1^2$).
Moreover, as it is clear from the Table~\ref{T:3},
the three fourth-order invariants built with the octupole
amplitudes ($C_3,\ C_4,\ C_5$)
and the square of the corresponding second-order invariant ($A_2^2$)
are not linearly independent of one another, and provide only two 
independent relations.

One can finally observe that the variables $X,\ Y,\ \eta$ and $\zeta $ 
always appear only in the combinations $(X^2 + Y^2)$, 
$(\eta^2 + \zeta^2)$.
As long as the expression of the potential energy does only depend
on the invariants up to the fourth order, 
the angles $\vartheta$ and $\varphi$ defined in the  Eq. (\ref{E:4.5})
are not subject to a restoring force.
The situation is more complicated for the three variables related to
the $\mu =\pm2$ components of $a^{\lambda}_\mu$. Also in this case,
however, it is possible to construct combinations of invariants that
contain only linear combinations of the squares of the variables 
$\xi$, $\gamma =\sqrt{5} \gamma_2 -\gamma_3$, and
$\gamma_0=\beta_2^2\gamma_2 +\sqrt{5} \beta_3^2 \gamma_3$.
Moreover, in this case, the first two of them do appear in the 
combination
$\beta_2\beta_3 \gamma^2 + 2\big( \beta_2^2+5\beta_3^2 \big) \xi^2$.
It is therefore possible to imaging a situation in which the potential
energy is independent also of the angle $\chi$ (although this is is not
a direct consequence of the model, as it is the case for the angles 
$\vartheta$ and $\varphi$).

\begin{acknowledgments}
We have the pleasure to thank
 Prof. F. Iachello and Prof. B.R. Mottelson for helpful discussions.
\end{acknowledgments}

\bibliography{biz_ref}

\end{document}